\documentclass[12pt,a4paper]{article}
\usepackage{cite,color,graphics,amsmath,epsfig,rotating}
\pdfoutput=1
   
\usepackage{latexsym}
\usepackage{amssymb}
\usepackage{slashed}
\usepackage{mathrsfs}
\usepackage{graphicx}
\usepackage{subfig}
\usepackage{psfrag}
\usepackage{url}
\DeclareMathAlphabet{\mathpzc}{OT1}{pzc}{m}{it}

\usepackage{hyperref}
\hypersetup{colorlinks=true,linktocpage,bookmarksopen,bookmarksnumbered,pdfstartview=FitH,breaklinks=true}
\ifx\pdfoutput\undefined

\usepackage[bookmarks]{hyperref}        
\usepackage{enumitem}
\else
\usepackage{hyperref}   
\fi

\usepackage{multicol}
\usepackage{color}
\usepackage{marginnote}
\usepackage{eurosym}

\def\micro{{\tt micrOMEGAs}}

\usepackage[pagewise,mathlines]{lineno}

\begin{document}
\begin{center}

{\bf Recasting direct detection limits within micrOMEGAs and implication for non-standard  Dark Matter scenarios.}

\vspace*{1cm}\renewcommand{\thefootnote}{\fnsymbol{footnote}}

{\large  G.~B\'elanger$^{1}$, 
A.~Mjallal$^{1}$, 
A.~Pukhov$^{2}$

\renewcommand{\thefootnote}{\arabic{footnote}}

\vspace*{1cm} 
{\normalsize \it 
$^1\,$ \href{http://lapth.cnrs.fr}{LAPTh}, Univ. Grenoble Alpes, USMB,CNRS, F-74940 Annecy, France\\[2mm]
$^2\,$\href{http://theory.sinp.msu.ru}{Skobeltsyn Institute of Nuclear Physics}, Moscow State University,\\ Moscow 119992, Russia\\[2mm]
}}

\vspace{1cm}

\begin{abstract}
 Direct detection experiments obtain  90\% upper limits on the  elastic scattering cross sections of dark matter with nucleons assuming  point-like interactions and standard astrophysical and cosmological parameters. In this paper we provide a recasting of the limits from XENON1T, PICO-60, CRESST-III and DarkSide-50 and include them in  micrOMEGAs. The code can then be used  to directly impose  constraints from these experiments on generic dark matter models under different assumptions about the DM  velocity distribution or on the  nucleus form factors. 
  Moreover, new limits on the elastic scattering cross sections can be obtained in the presence of a light t-channel mediator or of millicharged particles. 
  \end{abstract} 

\end{center}

\section{ Introduction}

Searches for dark matter(DM) through direct detection (DD) experiments have been pursued actively for
decades~\cite{Aprile:2018dbl,Aprile:2019dbj,Agnes:2018ves, Amole:2017dex, Amole:2019fdf, Petricca:2017zdp,
Xia:2018qgs,Agnese:2018gze}. None of  the experiments  with a good signal/background discrimination have found evidence for DM, thus could only set upper limits on the DM elastic scattering cross section on nucleons. 
For DM masses above roughly 3 GeV,  the best limits for spin-independent (SI) interactions are currently obtained by
XENON1T~\cite{Aprile:2018dbl,Aprile:2019xxb}. For lower masses, searches are more challenging and require a very low threshold for nuclear recoil energy, thus the limits are typically  much weaker.  Currently the best limits are obtained from DarkSide~\cite{Agnes:2018ves},  and CRESST~\cite{Abdelhameed:2019hmk}   and a series of projects are concentrating their efforts in improving the reach at or even below the GeV~\cite{Agnese:2015nto,Aprile:2019jmx} in particular by using DM scattering on electrons~\cite{Essig:2011nj,Agnese:2018col,Abramoff:2019dfb,Aguilar-Arevalo:2019wdi,Arnaud:2020svb,Aprile:2020tmw}.  For spin-dependent interactions on neutrons and protons, currently the best limits are obtained by XENON1T~\cite{Aprile:2019dbj} and PICO-60~\cite{Amole:2017dex, Amole:2019fdf} respectively.
Currently,  limits are generally interpreted in terms of DM elastic scattering on nucleons through a mediator with a mass much larger than the typical momentum exchange. Moreover they are 
 obtained assuming equal proton and neutron spin-independent  cross sections and  for a specific choice of
astrophysical parameters, notably that the DM velocity distribution is Maxwellian. 
 
Although traditional WIMP models feature  mediators at or above the electroweak scale (e.g., a Higgs, Z, a  new boson or a new coloured particle), new classes of DM models have relinquished the link with the electroweak scale thus  considerably extending the range of masses for both DM and mediators. In particular models with  a very light mediator have  been considered~\cite{Buckley:2009in,Bringmann:2016din,Kahlhoefer:2017ddj}. The  motivation for a light mediator  include the possibility to provide strong dark matter self-interactions and explain anomalies in galaxy clusters~\cite{Aarssen:2012fx,Tulin:2013teo,Kaplinghat:2015aga} as well as the possibility to enhance the direct detection signal in models with feebly coupled particles~\cite{Hambye:2018dpi}. 

While it is  straightforward for the experimental collaborations to obtain limits within a framework different than the default one chosen, the corresponding code is not publicly available. For example only PandaX~\cite{Xia:2018qgs,Cui:2017nnn} and more recently XENON1T~\cite{Aprile:2019xxb} have published limits obtained for both heavy and light
mediators. 
Our goal is precisely to provide a tool that allows to reinterpret the 90\% limits obtained by the experimental collaborations within their specific framework  and apply them  to a wider set of DM models and DM velocity distributions.  The code is developed as a module  of micrOMEGAs ~\cite{Belanger:2008sj,Belanger:2018mqt}. 
In this first version, a recast of the limits  from
XENON1T~\cite{Aprile:2018dbl}, DarkSide-50~\cite{Agnes:2018ves}, PICO-60~\cite{Amole:2019fdf} and CRESST-III~\cite{Abdelhameed:2019hmk}  are provided. These thus provide the best limits for  the cases of spin independent and spin dependent interactions in neutrons and protons for  DM masses above  1 GeV.  Based on  this recast, we  give typical examples on how the code can be used to set limits on new models. The models considered include the case of a light mediator, in particular a $Z'$,  as well as millicharged particles. Moreover the impact of  alternate velocity distributions is analysed.  Recasting of these limits as well as other recent  direct detection experiments are also included in DDCalc~\cite{Workgroup:2017lvb,Athron:2018ipf,Athron:2018hpc} and in SuperIso~\cite{Arbey:2018msw}. Note that both these recasting reproduce well the XENON1T exclusions for DM masses at the weak scale or above, however they feature significant differences for masses near the sensitivity threshold  when events are expected in the region at low nuclear recoil energy which is particularly challenging for experiments.
Our implementation provides a better match to XENON1T  in the case of light dark matter as will be described in the next section. Moreover since a heavy DM with a light mediator features a recoil energy distribution that resembles that of a light dark matter, in the sense that it peaks at smaller energies than the corresponding one for a heavy mediator, we expect a more reliable recast for the light mediator case. 
Considering the lack of complete information on the experimental data, for XENON1T we adopt a strategy which consists in tuning  the efficiency for nuclear recoils  in order to reproduce the SI  experimental  limit for  DM interactions  at all masses. We refer to this approach as 'inverse recasting'. 
Note that our approach can only be applied to the case where the DM signal is dominant at low recoil energy as will be discussed in Section~\ref{Xenon1Trecast}.  For  exotic signals with interactions at large recoil energy, for example the  ones studied in \cite{Aprile:2017aas, Angloher:2018fcs,Bozorgnia:2018jep}, our approach cannot be applied as it would  lead to  limits on the exclusion cross- section that are not severe enough. 
For other experiments we simply use the information provided in the publications to describe the detector efficiency and the background to reproduce the experimental limit.

The paper is organised as follows. After describing the formalism for the event rates in direct detection in section \ref{Experiments},   we describe our reconstruction
of  the XENON1T, DarkSide-50, PICO-60 and CRESST-III experimental limits on SI interactions    in Section 3 and SD ones in section 4. In section 5 we show how these recasts allow to obtain   limits in specific models involving  a   light mediator, a millicharge DM
as well as  generic DM velocity distribution. Section 6 contains our conclusions. 
All results obtained in our paper can be reproduced using the new micrOMEGAs functions described in  the Appendix \ref{Appendix}.

\section{Dark matter scattering on nuclei}
\label{Experiments}
We first review the standard formalism for obtaining the nuclear recoil energy distribution for DM scattering on nuclei, relevant for direct detection experiments. 
Since the velocity of DM particles is about  $v_0\approx 0.001c$, the maximum velocity of  the nucleus that recoils cannot exceed 2$v_0$.  Thus, the maximum transferred momentum in DM-nucleus collision is 
$q_{max}=2 v_0 M_A\approx 200$ MeV for a  nucleus mass $M_A \approx 100 {\rm GeV}$. At such low momentum transfer,  DM-nucleon interactions can be described by an effective Lagrangian  leading to constant matrix elements.  Moreover the amplitudes can be divided into  spin-dependent (SD) and spin-independent (SI) interactions which do not interfere. 
The DM-nuclei interactions are simply related to the DM-nucleon interactions after introducing a nucleus form factor which depends on the momentum transfer $q=\sqrt{2 M_A E}$ where $E$ is the nucleus recoil energy. 
The  energy  distribution of a recoil nuclei A   produced by SI interaction with DM  in a detector with total mass   $M_{det}$  and exposure time $T$  reads
\cite{Lewin:1995rx, Belanger:2008sj}
\begin{eqnarray}
\label{eq:SI}
   \frac{dN^{SI}_A}{dE} &=& \frac{2}{\pi}M_{det} T \frac{\rho_{\chi}}{M_\chi} I(E) 
  (\lambda_p Z + \lambda_n(A-Z))^2 F_A^2(q)
\end{eqnarray}  
where  $Z$  and $A$ are the  atomic number and mass of the detector material, $M_\chi$ is the DM mass, $\rho_{\chi}$  the DM local density,  and $\lambda_N$  are
DM-nucleon scattering amplitudes.  SI interactions are typically generated from effective scalar or vector interactions of DM with nucleons. For example for an effective scalar interaction of Majorana fermions with nucleons N, ${\cal L}=\lambda_N \bar\chi\chi \bar{\psi}_N{\psi}_N$,  the SI DM-nucleon cross section is given by
\begin{equation}
  \sigma_{\chi N}^{SI} = \frac{4} {\pi} \mu_{\chi N}^2\lambda_N^2\;\;,\;\;   N=n,p
\end{equation}
where   $\mu_{\chi N} = M_\chi M_N/(M_\chi+M_N)$ is  the   DM-nucleon reduced mass.  The event rate also depends on the nucleus form factor, $F_A(q)$ and on the velocity distribution through, 
\begin{equation}
     I(E) = \int \limits_{\sqrt{E M_A/(2\mu_{\chi A}^2)}   }^\infty \frac{f(v)}{v} dv\;\;, 
\end{equation}  
where  $f(v)$ is the DM velocity distribution in  the detector rest frame normalized such that
\begin{equation}
\label{normf}
   \int \limits_0^\infty f(v)dv =1 
\end{equation}
The recoil energy distributions for various DM masses are displayed in Fig.~\ref{35_200}-left.

In direct detection experiments  after analysing  the number of registered  events and estimating the  background, 
  limits   are set   on  $\sigma^{SI}_{\chi p}$ assuming  $\sigma^{SI}_{\chi p}=\sigma^{SI}_{\chi n}$.
All experiments also assume a value for the  DM local density near the Sun,   $\rho_{\chi}=0.3~{\rm GeV}/{\rm cm}^3$,  and a Maxwellian DM velocity distribution defined with the parameters
\begin{equation}
\label{velo_std}
 v_{Rot}=220~{\rm km/s}\;\;\;\; v_{esc}=544~{\rm km/s}\ \;\;\; v_{Earth}=232~{\rm km/s}
\end{equation} 
where  $v_{Rot}$, the rotation velocity of the Galaxy  and $v_{esc}$, the escape velocity in the galaxy, characterize the  DM velocity
distribution in the Milky Way  \cite{Lewin:1995rx}.  $v_{Earth}$ is  the velocity of the Earth in the galactic frame.

The energy  distribution  of  recoil  events  resulting from SD interactions of DM with nuclei in a  detector with mass $M_{det}$ and exposure time $T$  reads
\cite{Lewin:1995rx, Bednyakov:2006ux,  Belanger:2008sj, Klos:2013rwa}
\begin{equation}
   \frac{dN^{SD}_A}{dE} = {\mathcal M}_{det} T \frac{\rho_{\chi}}{M_\chi} I(E)
                  \frac{8}{2 J_A+1}(S_{00}(q)(\xi_p+\xi_n)^2 +S_{01}(q)(\xi_p^2-\xi_n^2)+
S_{11}(q)(\xi_p-\xi_n)^2)
\label{eq:SD}
\end{equation}  
where  $J_A$ represents the  spin  of the  detector material,  $\xi_{p,n}$ are
 the DM-nucleon amplitudes normalized such that   
\begin{equation}
  \sigma_{\chi {N}}^{SD} = \frac{12}{\pi}\mu_{\chi{N}}^2\xi_{N}^2\;\;.
\end{equation}
  For example,  an effective axial-vector interaction of Majorana fermions with nucleons, ${\cal L}=\xi_N \bar\chi \gamma_\mu\gamma_ 5\chi \bar{\psi}_N \gamma^\mu\gamma_ 5{\psi}_N$ will lead to the above cross section while for Dirac fermions, the same cross section is obtained for a Lagrangian defined with  $\xi_N \rightarrow  2 \xi_N$.
 $S_{ij}(q)$ are the nucleus  SD form factors. Calculations or these form factors within nuclear models are reviewed in ~\cite{Bednyakov:2006ux} and more recent calculations are available in  ~\cite{Klos:2013rwa}.  Another set of form factors is currently used by experimental collaborations, these form factors, $F_{44}^{ab}$, are defined in the  effective field theory approach in Ref.~\cite{Fitzpatrick:2012ix}, they are expressed as
 \begin{equation}
   F_{44}^{ab}= \frac{J_A(J_A+1)}{12}(F_{\Sigma^{'}}^{ab}+F_{\Sigma^{''}}^{ab})
   \label{eq:sdff_fitzpatrick}
 \end{equation}
 and analytical expressions for $F_{\Sigma^{'}}^{ab}, F_{\Sigma^{''}}^{ab}$ can be found in the Appendix of Ref.~\cite{Fitzpatrick:2012ix}. Simple expressions allow to relate these form factors with those in Eq.~\ref{eq:SD}. 
 \begin{eqnarray}
\nonumber
F_{44}^{pp}(q)&=&\frac{\pi}{4(2J_A+1)} \left( S_{00}(q)+S_{11}(q)+ S_{01}(q)\right) \; , \\
F_{44}^{nn}(q)&=&\frac{\pi}{4(2J_A+1)} \left( S_{00}(q)+S_{11}(q) - S_{01}(q)\right) \nonumber\\
 F_{44}^{pn}(q)&=&F_{44}^{np}(q)= \frac{\pi}{4(2J_A+1)} \left( S_{00}(q) - S_{11}(q)\right).
 \end{eqnarray}
  
 \section{Spin-independent interactions: recasting experimental exclusions}

\subsection{XENON1T.} 
\label{Xenon1Trecast}

To repeat  exactly the XENON1T analysis would require detailed information on events distribution,
background estimation, and  the use of nuisance parameters for all points of event space  characterized by scintillation signals  $cS1,cS2b$  and interaction positions $Z$ and $R$ ~\cite{Aprile:2018dbl}.  
Lacking this detailed information we propose  instead to reconstruct an effective  efficiency  by  using the 90\% exclusion  cross section
obtained by XENON1T from their complete analysis, this will then be validated by comparing with the XENON1T upper
limit as will be explained below. This approach can be considered as a simplified version of the XENON1T analysis where some cuts in $cS1,cS2b$
space are applied to increase the signal/background ratio.
Our  simplified approach  relies on the observation that in some of the subspaces where XENON1T reported signal and background best-fit values, XENON1T can be considered as a low background experiment. 
Specifically we will use  the reference detector mass of 0.9t. In this subspace  illustrated in Fig.2 and  Fig.3  of Ref.~\cite{Aprile:2018dbl} both 
the electromagnetic and neutron background  are suppressed.  In this region, XENON1T reports two detected events and an estimated background, $n_B=1.62$ events. 
 

We first have to check that relying only on partial data is a reasonable assumption to approximately reproduce the upper limit obtained in the full analysis.
For this purpose
we use the data for  the best-fit point presented in Table 1 of  \cite{Aprile:2018dbl}.
For this reference point  the DM mass, $M_R$, the   cross section, $\sigma_R$,  and the  expected number of signal events after applying cuts, $n_R$, are given by 
\begin{equation}
\label{best-fit}
  M_R=200GeV\;\;\;\;\;   \sigma_R=4.7\times 10^{-47} {\rm cm}^2\;\;\;\;\; n_R=1.16\;\;\;,
\end{equation}
Using the  Feldman-Cousins formula we can easily estimate the cross section required for a 90\% exclusion, we find 
 $\sigma= 1.65\times10^{-46}{\rm cm}^2$, a value close to the one obtained by XENON1T,  $\sigma=1.73\times 10^{-46}{\rm cm}^2$.
Thus we conclude that the XENON1T data obtained after  imposing cuts is suitable for obtaining upper  limits
on the DM-nucleon cross section.

In general to recast the result of a DD experiment 
 while lacking the full information on signal events, cuts, backgrounds and the associated uncertainties, one needs at least 
to know the detection efficiency $p(E)$ and the  background
distribution  as  function of the nucleus recoil energy after cuts.  
The efficiency of XENON1T, which we denote $p_{Xe}$,  is shown in Fig.1 of Ref.~\cite{Aprile:2018dbl}. We use the efficiency of the second science Run, SR1. However, this efficiency
does not include the effect of $cS1,cS2b$ cuts. Indeed the number of signal events for the best-fit point obtained with this efficiency and for the full detector mass 1.3t, is $n=3.56$ which 
  corresponds to the number of DM signals before cuts cited in  Table I \cite{Aprile:2018dbl}. The same table shows that this number is reduced by a factor 1.7/3.56 after cuts.~\footnote{In DDCalc~\cite{Athron:2018hpc}, an
  overall reduction factor  1.7/3.56  is applied to $p_{Xe}$ to take into account the effect of cS1,cS2b cuts, thus the recast of the 90\% excluded cross section for light DM is more than a factor 2 above that of XENON1T, see Fig.13 in Ref.~\cite{Athron:2018hpc}.}
   Moreover we note that using the efficiency $p_{Xe}$ for the excluded signal for a DM of 6GeV ($\sigma =2.8 \times 10^{-8}{\rm pb}$) we obtain only 1.3 events, a number insufficient for a 90\% exclusion. Thus we choose not to use directly $p_{Xe}$ and instead
 propose   to reconstruct an effective  efficiency  by  using the 90\% exclusion  cross section
obtained by XENON1T from their complete analysis, this will then be validated by comparing with the XENON1T upper limit.

For a  low background experiment it is reasonable to use non-binned likelihood
\begin{eqnarray}
\label{LLforEvents}
\nonumber
  &&   L(p,\frac{dN_{\chi}(M_{\chi},\sigma)}{dE}) = 
 e^{- {\cal L} \int \limits_0^{E_{max}}
\left( p(E)\left(\frac{dN_{\chi}(M_{\chi},\sigma )}{dE} + b(E)\right)+b_\gamma(E) \right)
dE}  \prod\limits_{k\in events}\left(b_k +s_k \right)\\
\label{LLforEvents2}
&=&\left[e^{- {\cal L} \int \limits_0^{E_{max}}
(p(E) b(E) +b_\gamma(E))dE} \prod\limits_{k\in events} {b_k}\right]
 e^{- {\cal L} \int\limits_0^{E_{max}}
p(E)\frac{dN_{\chi}(M_{\chi},\sigma )}{dE} 
dE}  \prod\limits_{k\in events}\left(1+ \frac{s_k}{b_k} \right)
\end{eqnarray}
where  $dN_{\chi}/dE$ is the nuclei recoil energy distribution corresponding to the scattering of a   DM of mass $M_\chi$ with a cross section $\sigma$,  $p(E)$ is the efficiency for  the detection of  signal events,    ${\cal L}$ is the  exposure,  $b(E)$ and $b_\gamma(E)$ are the neutron and electromagnetic  background distribution, and  $s_k= \frac{dN_{\chi}(M_{\chi},\sigma)}{dE_k}$ and $ b_k$ are  the  signal and  background  probability distribution function (p.d.f.)  for each detected event. 
Using a Bayesian approach with flat priors, we determine the credible interval for the cross section  [$0,\sigma_{ex}$] corresponding to a fraction $1- \alpha$ of the posterior probability where, 

\begin{equation}
\label{LLratioB}
   \alpha(\sigma_{ex})=\frac{ \int_{\sigma_{ex}}^\infty    L(p,   {dN(M_{\chi}, \sigma )}/{dE})d\sigma } {
\int_0^{\infty}  L(p,  { dN(M_{\chi},  \sigma)}/{dE} )
d\sigma}    
\end{equation}

 Note that when nuisance parameters are taken into account when estimating  the background, one has to integrate both the numerator and denominator in  Eq.\ref{LLratioB}  with some prior. 
In the following we ignore such nuisance parameters,   thus  the term enclosed in squared brackets in
Eq.~\ref{LLforEvents2} cancels out. In this approximation, the background contributes only via the ratio ${s_k}/{b_k}$ in Eq. ~\ref{LLforEvents2}.
From Fig.~3 in Ref.~\cite{Aprile:2018dbl}, we deduce that the two events detected  by XENON1T correspond to
\{cS1,cS2b\} coordinates \{17,400\} and \{50,1300\} from which we estimate $E_r =  12 , 33~{\rm keV}$ and $s_k/b_k=$ 0.7, 0.2 respectively.

For an approximate recast of XENON1T for all masses,  in particular for low DM masses and small recoil energies, we
consider
Eq.\ref{LLratioB} as an equation for the efficiency $p(E)$ that has to be satisfied for all masses in the interval [6-1000] GeV, for 
$\alpha=0.1$ and the cross sections $\sigma=\sigma^{90}(M_{\chi})$ obtained by XENON1T for SI interactions.  

\begin{figure}[htb]
\centering
\includegraphics[scale=0.59]{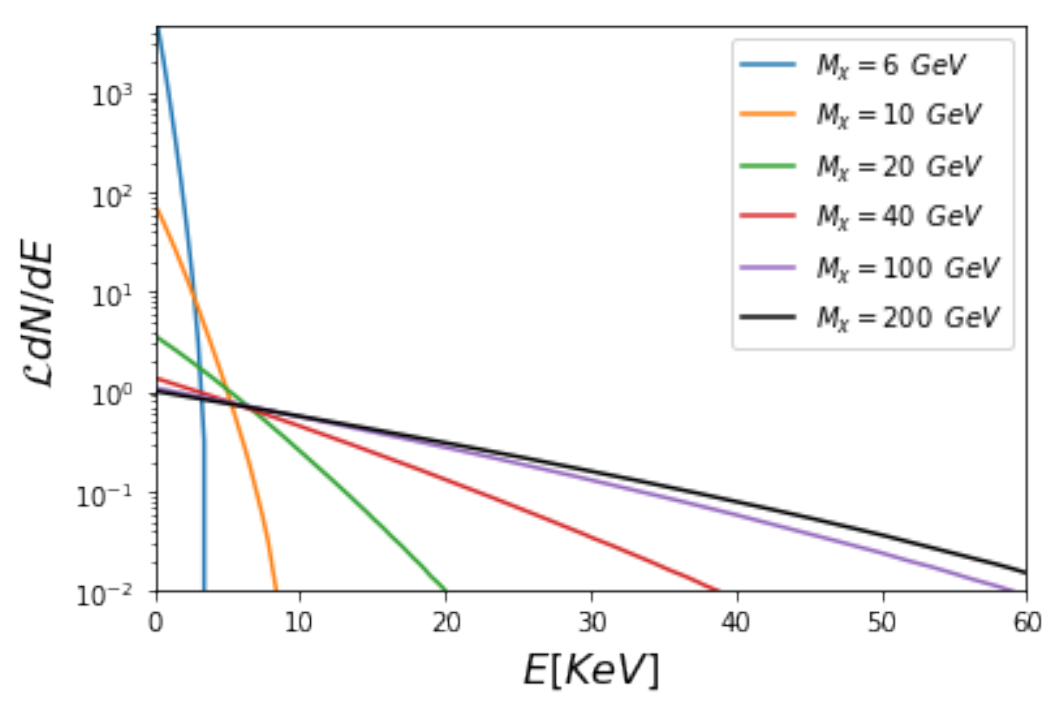}
\includegraphics[scale=0.45]{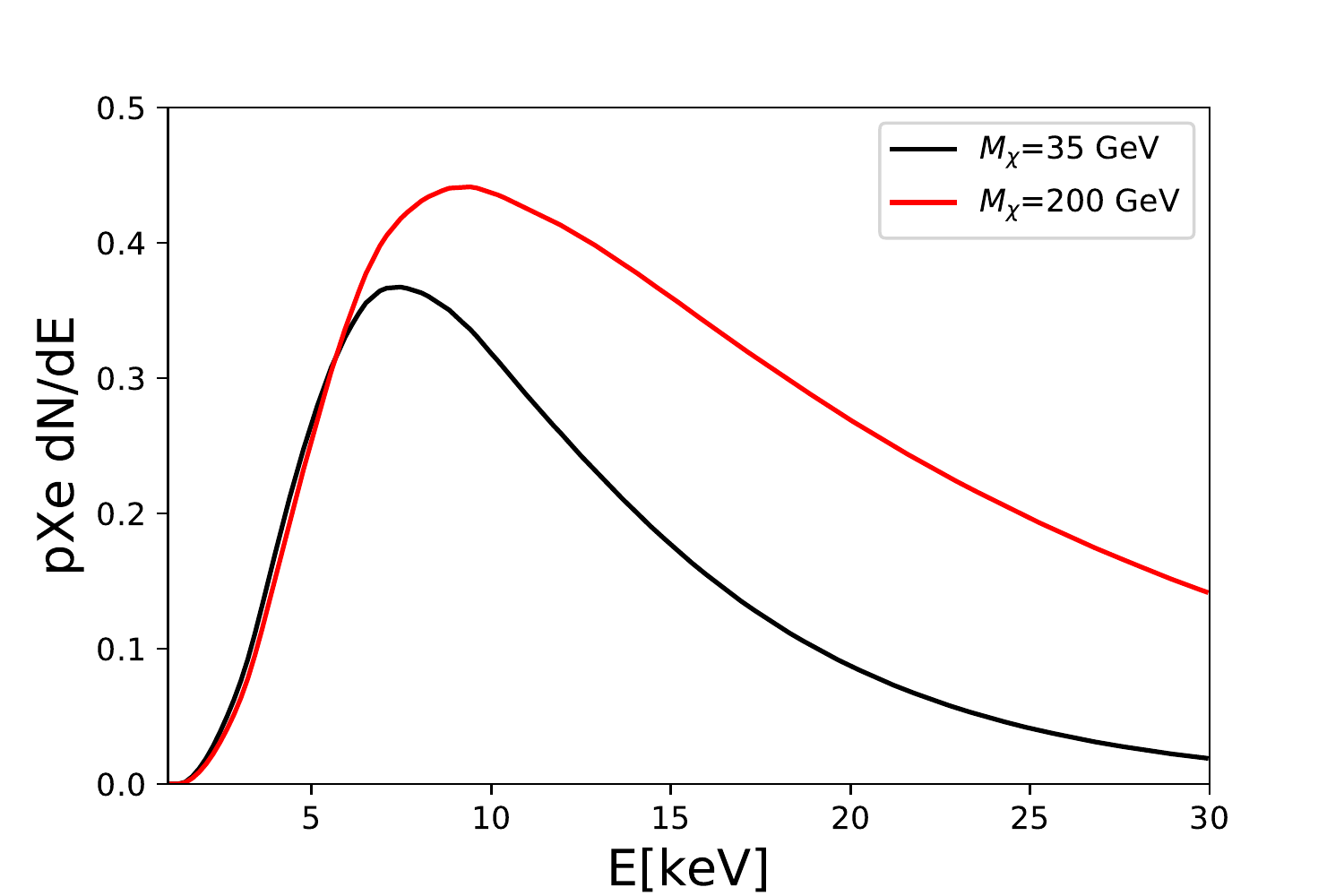}
\vspace{-.3cm}
\caption{Left: Predictions for  the  recoil energy  distribution of Xenon nuclei  for an exposure ${\cal L}=279\times900$ kg$\cdot$days
and  $\sigma^{90}=248.6,5.39,0.566,0.448,0.912,1.71\times 10^{-46}{\rm cm}^2$ for $M_\chi=6,10,20,40,100,200$~GeV respectively.
Right: Recoil energy distributions convoluted with the XENON1T acceptance $p_{Xe}(E)$ for   $M_\chi=35(200)$~GeV and
 $\sigma^{90}=4.71 (17.1)\times 10^{-47} {\rm cm}^2$.
  }
\label{35_200}
\end{figure}

 At first approximation we neglect  the last term  in Eq.\ref{LLforEvents}, thus assuming that  there is some effective subspace in the S1/S2 parameter space where no events were detected and which can be used to reproduce the exclusion cross section. \footnote{Note that XENON1T uses a frequentist approach in their full analysis. In the subspace where no events were detected choosing a different statistical approach as we do here should not have a large impact on the 90\% excluded cross-section presented in
Section~\ref{Applications}. }
 This approximation is motivated by a  comparison of  the recoil energy
distributions corresponding to DM masses of 35 GeV and 200 GeV shown in  Fig.\ref{35_200}-right  for $\sigma^{90}(M_{\chi})$ of XENON1T and using $p_{Xe}$.  Clearly these two signals  practically coincide  for low energies  while the signal for 200 GeV becomes much larger  for
$E\gtrsim 8~{\rm keV}$. 
 These two signals leading to the same level of exclusion might indicate that the events with large recoil energies do not contribute significantly to the 90\% exclusion.
It is indeed expected that for a low background experiment the  region where events are found (here at energies above 12 keV) does not contribute significantly to the exclusion.
Our approach can be considered as a simplified version of the XENON1T analysis  where some cuts in
cS1/cS2b space are applied to increase the signal/background ratio.

We denote $p_{eff}^0(E)$  the effective detector efficiency after all cuts assuming no events were detected. Eq.\ref{LLratioB} leads to an integral equation for  $p_{eff}^0(E)$ for all masses  
in the range $6~{\rm GeV}<M_{\chi }<1000~ {\rm GeV}$,
 \begin{equation}
\label{PoissonNew}
 {\cal L} \int p_{eff}^0(E)\frac{dN(M_{\chi},\sigma^{90}(M_{\chi})}{dE} dE =
-\log(\alpha)=log(10).
\end{equation} 
Equation \ref{PoissonNew} is a  Fredholm equations of the first kind. The solution of such
equations is not stable  and leads to large oscillations in  $p_{eff}(E)$. To smooth out these oscillations, rather than solving it directly we  minimize the functional
\begin{equation}   
  J(p_{eff})= \max_{M_{\chi}}\left( \left| {\cal L}  \int
p_{eff}(E)\frac{dN(M_{\chi},\sigma^{90})}{ dE} dE +log(\alpha) \right| + \kappa
 \int \left|\frac{d^2}{dE^2}p_{eff}(E)\right|^2 dE   \right)
\label{regEq}
\end{equation}
 with respect to the function $p_{eff}(E)$.   The minimization covers all masses in the  interval considered, thus allowing to obtain a good agreement for each mass. Note that the term with  $\kappa$  damps 
oscillations only if $\kappa$ is large enough, while it spoils the solution to 
Eq.\ref{PoissonNew} when $\kappa$ becomes too large. The goal is therefore to find the minimal $\kappa$ which allows to obtain a  solution without oscillations.
 To find the minimum of $J(p_{eff})$ we tabulate $p_{eff}(E)$ on a grid which 
extends from $E_0$ to some $E_{max}$ with a  1keV step size. The acceptance $p_{eff}(E)$ vanishes for $E\le E_0$, where $E_0$,
  the detection threshold,  is taken as a free parameter as well. It is found to be $E_0=1$keV. The 
values of the  function $p_{eff}(E)$  at each point on the grid except the first one are also free parameters and we impose the condition that $p_{eff}(E)\ge 0$ . Between grid points we use a cubic polynomial interpolation. 

\begin{figure}[htb]
\centering
\includegraphics[scale=0.39]{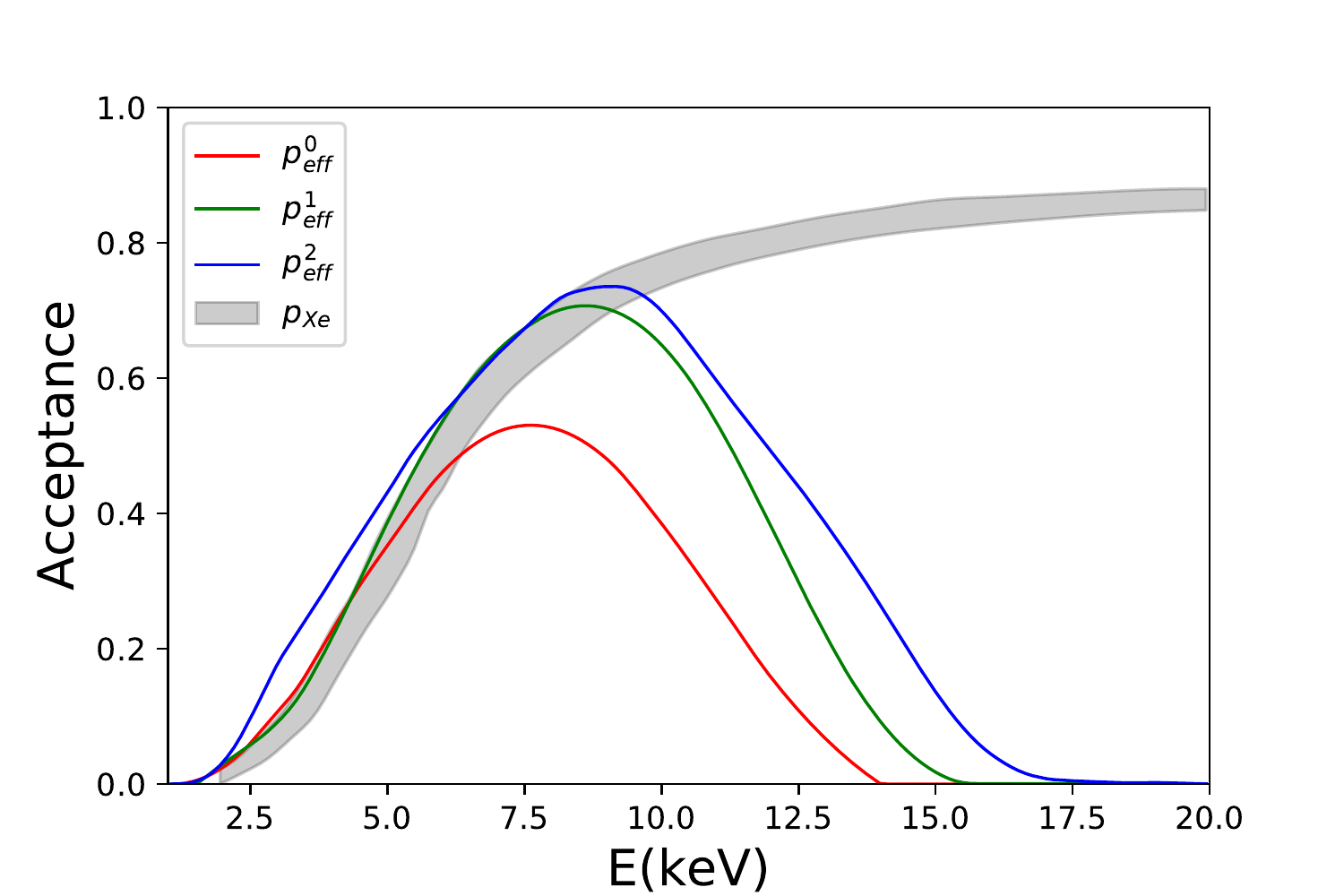}
\includegraphics[scale=0.59]{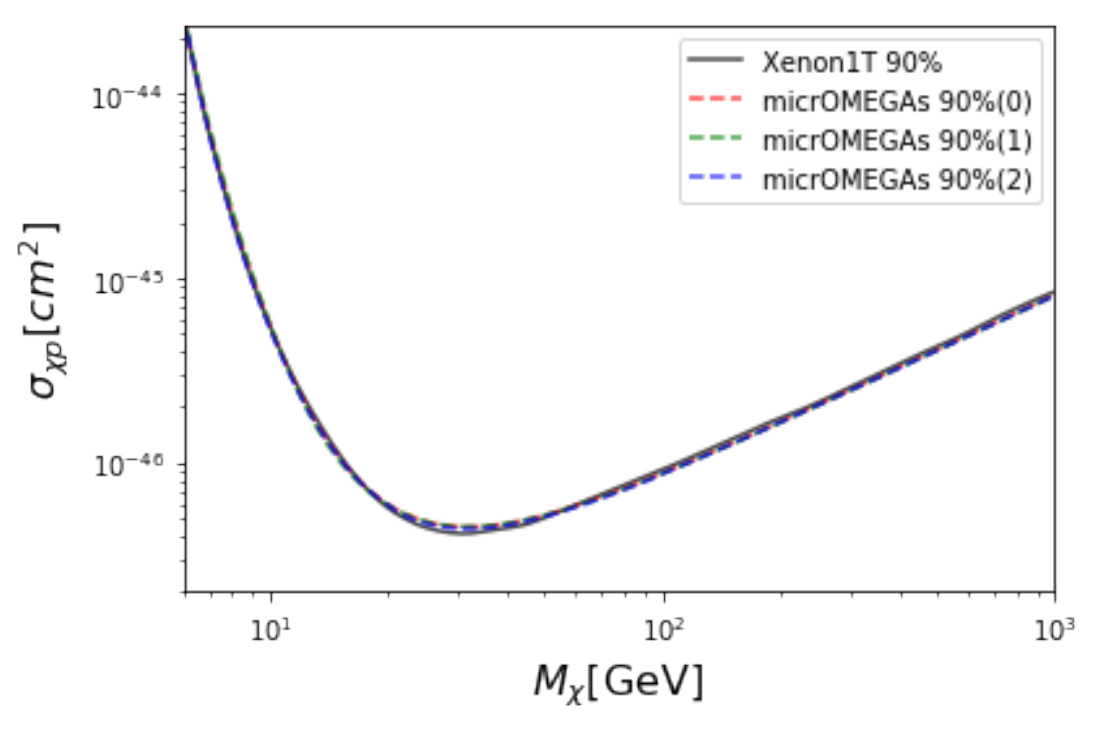}
\vspace{-.3cm}
\caption{ Left:  The reconstructed acceptances  $p_{eff}^0(E)$ , $p_{eff}^1(E)$  and $p_{eff}^2(E)$ compared to the XENON1T   total ($p_{Xe}$) efficiency corresponding to SR1  from Fig.1 in Ref.~\cite{Aprile:2018dbl} (grey band).  Right : The 90\% excluded cross section for SI interactions obtained with $p_{eff}^0(E)$ (red-dash), $p_{eff}^1(E)$ (green-dash) and 
$p_{eff}^2(E)$ (blue-dash)  as compared with XENON1T (black).
}.  
\label{p0Znew}
\end{figure}

The solution $p_{eff}^0$ is shown in Fig.\ref{p0Znew}-left and is compared to $p_{Xe}$ and $p_{Xe}^{D}$, the XENON1T  total and detection efficiencies respectively in Ref.~\cite{Aprile:2018dbl}. For the total efficiency $p_{Xe}$  corresponds to the second science run (SR1). We find that  $p_{eff}^0$ nearly vanishes 
at the recoil energy of  the  detected event with the smallest recoil energy.  This condition was not imposed in advance and testifies 
of  the  validity of our assumption. Indeed if we had found a non-negligible  $p_{eff}^0$  in the region where events are detected,  we would not be able to conclude that the results of  XENON1T which were obtained using the full  events space and including all 735 observed events in the likelihood, 
can be reproduced  in the  zero-event approximation.  
\footnote{
For instance, applying the same method to XENON100, leads to  an efficiency that is not negligible in the region where events were detected, thus 
the solution  $p^0_{eff}$  can not be blindly applied for any experiment.} 
The fact that $p_{eff}^0$ and $p_{Xe}$ are comparable (allowing for the uncertainty in the XENON1T efficiency) at low energies is also consistent with the observation that for light DM (say 10GeV) for which the signal is concentrated at low recoil energies, the signal is located in a region without electronic recoil  background events as can be seen by comparing Fig. 8 in ~\cite{Aprile:2019dme} and Fig.3 in ~\cite{Aprile:2018dbl}.  We observe also that close to threshold, the efficiency $p_{eff}^{0}$ is slightly  larger than $p_{Xe}(E)$, 
this is probably  related to the fact that in this region the efficiencies rise sharply hence have  larger uncertainties.

Some comments are in order. First, it is well known that the Bayesian credible intervals significantly  depend on priors in the no-event case,  for example, with the flat prior used the 90\% exclusion corresponds to $\approx 2.3$ signal events while for 
Jeffreys prior it corresponds to $\approx 1.3$ events. Changing the prior in Eq.\ref{LLratioB} would therefore rescale the efficiency obtained, however there would be no impact on  the excluded cross section as the same prior is used to fit the efficiency and to calculate the exclusion. 
 Second, we chose Bayesian statistics over the frequentist approach adopted by XENON1T  because we have no information about the background distribution.  
In case of  zero event and background, the upper limit for exclusion  depends only on  the number of predicted signal events, this in turn depends on the choice of the statistical method. Thus all approaches, whether Bayesian or frequentist,  will lead to  the same reconstructed efficiency  $p_{eff}^{0}$ up to an overall scaling factor. This means that  in the framework of inverse recasting, one will reproduce the same result for the  90\% upper limit on the DM scattering cross-section. 
Moreover in the low recoil energy interval (1-14keV)  we estimate the background to be low, only 0.35 events, assuming a simple counting experiment with Feldman-Cousins statistics we can estimate that the number of events required for a 90\% exclusion with such background changes from 2.44 to 2.08, hence a 15\% correction. Thus including background would  only slightly modify the   efficiency   $p_{eff}^{0}$, without  affecting  the exclusion limit.

Following the same procedure,  we also derive the efficiency  $p_{eff}^1$  taking into account  one
 detected event in the subsample at $E_r \approx 12 {\rm keV}$. The corresponding solution for the efficiency is shown in  
Fig.\ref{p0Znew}.     At last we find the efficiency $p_{eff}^2$ which allows to
reproduce the XENON1T exclusion curve using  the extended   optimum interval method by Yellin \cite{Yellin:2008da}. In
this case we assume  that the background is uniformly distributed in the interval [1-50] keV.  
Note that  for the best-fit signal  of XENON1T, $p_{eff}^2$ leads to  1.04 events, only  slightly
smaller than  $n_R$ in Eq.\ref{best-fit}.

The recasted  90\% excluded cross sections obtained with $p_{eff}^{0,1,2}$  are displayed  in
Fig.\ref{p0Znew}-right. 
All  allow to reproduce the XENON1T exclusion within 10\%. The largest difference is found 
near   $M_{\chi} \approx 35GeV$, see Fig.\ref{p0Znew}. We have compared the results for the three different recasts for the applications in Section~\ref{Applications} and found no significant difference. 
 Note that  we did some approximations,  for instance, we use only a sub-space of the full analysis where background is small, 
we ignored  background uncertainty as well as the uncertainty in the energy of detected events. 
However the error introduced by these approximations are compensated by the fact that we fit the exclusion curve when solving for $p_{eff}$.
 In section~\ref{SDinteractions}, we will see that despite these approximations, our recast works well for the  slightly different recoil spectra that are expected for SD interactions.

  To conclude this section we emphasize  that our method of reconstruction of the XENON1T efficiency neglects the contribution from large recoil energies.
  Thus this  recast  cannot be used for models which produce a recoil energy spectrum dominant at large energies such as can be obtained with effective operators or with inelastic scattering~\cite{Aprile:2017aas, Angloher:2018fcs,Bozorgnia:2018jep}, it would lead to too conservative limits.

\subsection{DarkSide-50}
The DarkSide-50  (DS-50) collaboration \cite{Agnes:2018ves}  provides  the  basic experimental  data to allow   to reproduce the experimental
results using a standard procedure.  In particular,   the distribution for the  number of  ionizations $n_{e^-}$  in the  Argon detector  for an  exposure  ${\cal L}=6786 ~{\rm kg\cdot days}$ together with an estimation of the background  and the ionization quenching  are given. We use  the numerical tables for the data and background provided  by the DarkSide collaboration. We are thus able to construct a likelihood based 
on the Poisson formula

\begin{equation}
L=\prod_i \frac{(B_i+S_i)^{n_i}}{n_i!} \exp^{-(B_i+S_i)}
\label{eq:DS-50LL}
\end{equation}
where $B_i$ and $S_i$ are the number of background and signal events in the $i^{th}$ bin  where the bins are defined for the distribution of the number of ionizations.
 Following  DS-50 analysis,  we do not include the  bins  $n_{e^-}<4$  in the likelihood. For $ 4\le n_{e^-}<7$    
there is a large difference between the  data and the estimated
background, hence, following the DS-50 analysis, we treat the additional background as a nuisance parameter when constructing the likelihood function~\cite{Cowan:2010js}. 
It means that we include the contribution of the  $4\le n_{e^-}<7$  only if  the DM signal plus known background is
larger than the experimental data.

The average number of ionizations is determined by quenching.
The ionization quenching  depends on the recoil energy and  suffers from  a large uncertainty~\cite{Bezrukov:2010qa}. We use the minimal quenching. \footnote{There is an alternative estimation of the uncertainty on the number of ionizations ~\cite{Bernreuther:2019pfb}, here we rather use the quenching adopted by DS-50.}
We have checked that making a linear interpolation between the  minimal and maximal values of the quenching for each energy and treating  the parameter of interpolation  as a nuisance parameter leads to very similar results. Moreover, the distribution of the number of ionizations around the average is not known. The assumption made for describing this distribution is essential for light DM, since one can find events with  $n_{e^-}\ge 4$ that arise from the tail of the distribution.

DS-50 considers two cases, first  a binomial distribution for the  number of ionizations
where  the average number of ionizations is determined by quenching,  while the maximal number of ionizations is  determined by the minimal energy 
needed for one ionization. The minimal energy is set  to $E_1=19.5$ eV.
Second  a $\delta$-like distribution is considered where the actual number of events  equals the average one. 
 The key feature of the  binomial  distribution is that it allows to naturally implement an energy threshold which cannot be done with the widely used Poisson distribution.  
However  the Poisson distribution  can be generalized in order to take into account an energy threshold, thus we use the following distribution for the number of ionized electron :
\begin{equation}
   p(n) = \frac{C(\bar{n},E_1)}{n!} \left(\bar{n}(1-n\frac{E_1}{E})\right)^n
   \label{eq:poisson}        
\end{equation}
 Here $C$ is defined by the normalization condition, $E$ is the energy of an atom after DM recoils, $E_1$ is the minimal ionization energy. The value of  $\bar{n}$ is chosen 
 in such a way to reproduce the quenching given by  DS-50~\cite{Agnes:2018ves}.  This generalized Poisson distribution  has a  tail that  decreases faster than the binomial distribution and  thus  provides a more conservative limit. To compute the signal in each bin, $S_i$, and construct the likelihood  in Eq.~\ref{eq:DS-50LL}, we convolute the predicted recoil energy distribution with the generalized
Poisson distribution  for the number of ionized electron, Eq.~\ref{eq:poisson}, 
as well as with a gaussian with  20\% resolution.
The latter is used to describe the uncertainty introduced when DS-50 reconstructs the number of ionizations from their counts of
photons.

As the DS-50 collaboration, we use the profile likelihood ratio, $\lambda$, to  calculate the confidence level for excluding a model \cite{Cowan:2010js}

\begin{equation}
   C.L. = 1-\frac{1}{2} \int \limits_{-2\log(\lambda)}^{\infty} \frac{dx}{\sqrt{2\pi x}} e^{-\frac{x}{2}}
\end{equation}

Following the procedure of DS-50, we have implemented a nuisance parameter for the background which is represented by an overall factor of $\pm 15\%$. \footnote{Note that introducing an overall uncertainty factor does not improve the rather poor global $\chi^2/N$ which we estimate to be 2.8. }

\begin{figure}[htb]
\centering
\includegraphics[scale=0.56]{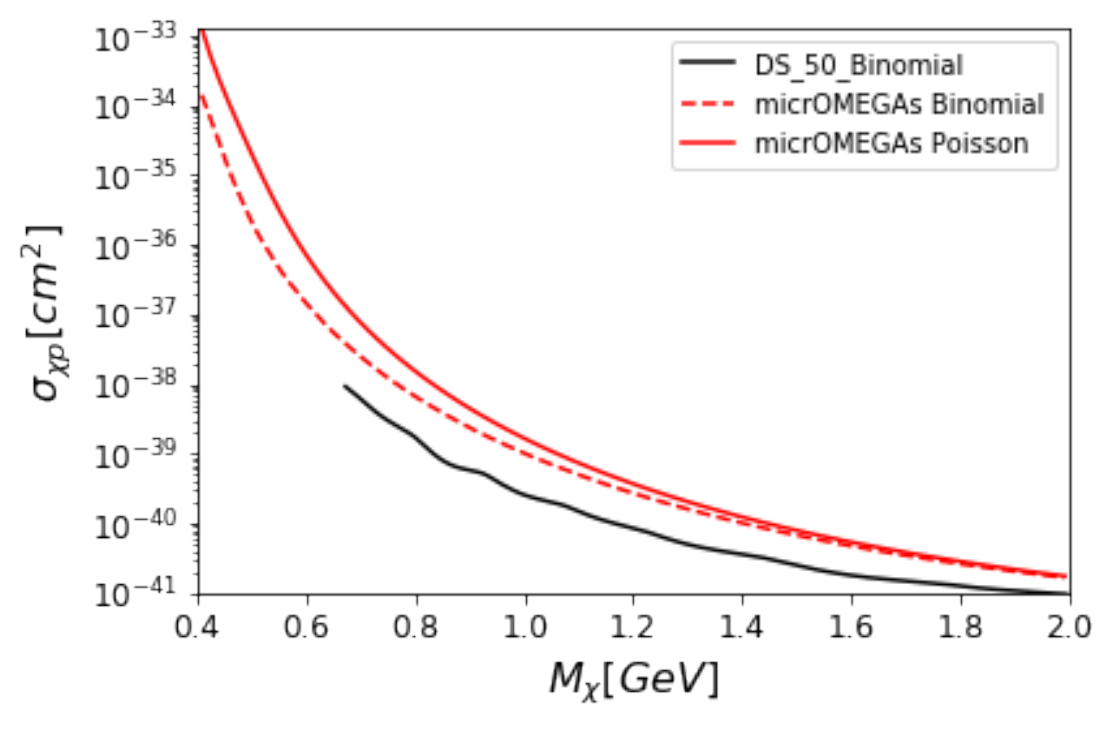}
\includegraphics[scale=0.56]{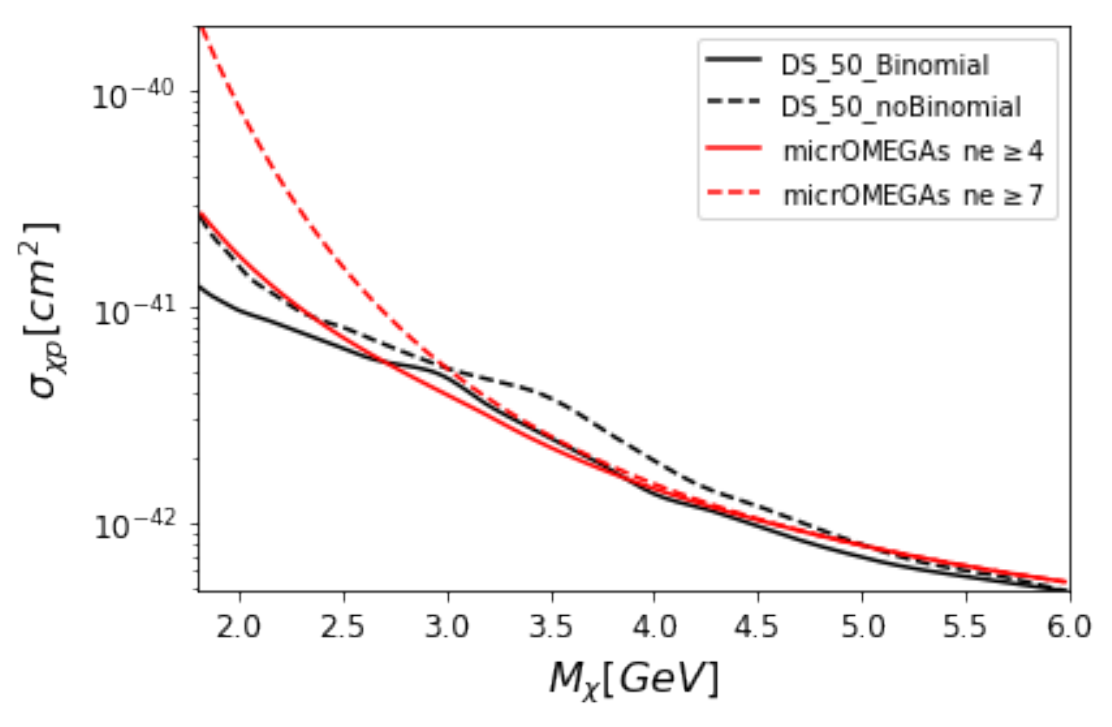}
\vspace{-.2cm}
\caption{Comparison between the 90\% excluded SI cross section on protons from DS-50 with a binomial distribution (full black)  and from  micrOMEGAs with the default option, a Poisson distribution using all bins with $n_{e-} \ge 4$ (full red). The left panel also shows  the exclusion  for light masses when using a binomial distribution in micrOMEGAs (red dash). The right panel shows the difference in the exclusion from DS-50 with (full) and without (dash) the binomial distribution as well as the impact of using only the bins with $n_{e-} \ge 7$ within micrOMEGAs (red dash).   }
\label{DS-reconstruction}
\end{figure}

For  $M_{\chi}>1.8GeV$ both the binomial and improved Poisson distributions  lead to the same exclusion. In Fig.\ref{DS-reconstruction} (right) we compare  our reconstructions of the 90\% excluded cross section
with the DS-50 exclusion. Note that DS-50 uses two different likelihoods, one using bins $n_{e^-}\ge 4$ for masses $M_{\chi}<2.9$~GeV and one using only the bins $n_{e^-}\ge 7$ for higher masses. 
Rather than splitting our analysis for different mass range and in order to have a smooth exclusion, we  take into account all bins $n_{e^-}\ge 4$ for the whole DM mass range. 
We still reproduce well the  DS-50 exclusion  for masses $M_{\chi}>3.5$~GeV, since in this region the contribution from higher bins dominate.  
Around $M_{\chi}\approx 3$~GeV our exclusion is stronger since  the bins $4\le n_{e^-}< 7$ give an important contribution to the likelihood. 
Finally our exclusion is more conservative at lower masses,  by about 50\%(200\%)   for $M_{\chi}\approx 1.8
(0.65)$~GeV.

In Fig.\ref{DS-reconstruction}-left we compare  the exclusion cross section obtained by DS-50 assuming a binomial distribution with the ones reconstructed by micrOMEGAs 
for both the  binomial and generalized Poisson distributions for  $M_\chi <2$~GeV, here we use all bins $n_{e^-}\ge 4$.
Note however that for masses 
below 1.8 GeV there are  large uncertainties in  the DS-50 exclusion  depending on the choice of quenching  model~\cite{Bezrukov:2010qa}.

\subsection{PICO-60}
\label{sec:pico}

PICO~\cite{Amole:2017dex, Amole:2019fdf}  is a  Bubble Chamber experiment which  uses $C_3F_8$, with 1167 kg-day exposure at a thermodynamic threshold of 3.3 keV and     
1404 kg$\cdot$ days at  2.45 keV.  After the acoustic parameter cut, PICO reports 3 candidate events for 
the second run while no events were detected in the first run~\cite{Amole:2017dex}. A combined analysis of both runs which includes a new efficiency for the first run was published  in Ref.~ \cite{Amole:2019fdf}.

\begin{figure}[htb]
\centering
\includegraphics[scale=0.65]{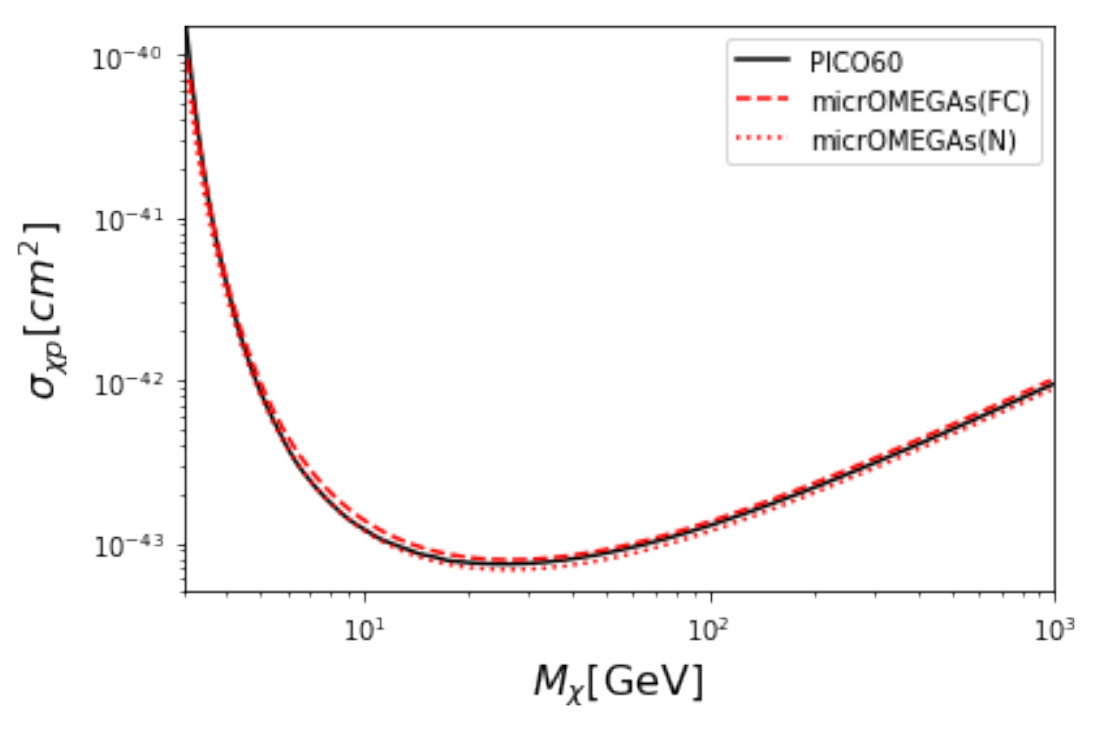}
\vspace{-.3cm}
\caption{ Comparison of PICO SI  90\% excluded cross section with our reconstructions based on
Felman-Cousins (FC) and Neyman methods(N).}
\label{PICO-reconstruction}
\end{figure}

  To reconstruct  the PICO-60 exclusion curve for SI interactions, we assume  the  central value of the acceptance shown in Fig. 3 of   Ref.~\cite{Amole:2019fdf} for each run. 
 In our statistical analysis we compare the total number of expected events for both runs combined with the total number of detected events. We estimate  the total background for both runs as $B=1.47$ events assuming that the ratio of   single to multiple bubble events caused by neutrons is 1/4.   
 We use two statistical  methods based on    Feldman-Cousins ~\cite{Feldman:1997qc} and   Neyman with one-side belt with the confidence level,
\begin{equation}
   C.L.=  \sum_{n=0}^{3} \frac{(S+B)^n}{n!} e^{-(S+B)}
\end{equation}
where $S$ is the number of predicted events caused by DM, and $B$ is the  expected background. 
In both cases we reproduce the 90\% exclusion for the SI cross section of PICO-60~\cite{Amole:2019fdf}   within 10\%, see Fig.\ref{PICO-reconstruction}.  
For the applications in the following sections we will use the  recasting based on  Feldman-Cousins.
The main result of PICO-60 however concerns  limits on the DM-proton SD cross section, this result will be discussed in Section~\ref{SDinteractions}.

\subsection{CRESST}

The CRESST-III detector uses $CaWO_4$ and the limits obtained correspond to data collected with a total exposure  of 5.594 kg$\cdot$days or 3.64 kg$\cdot$days after cuts~\cite{Abdelhameed:2019hmk}. 
In this experiment, the background is not estimated and the Optimum Interval method of Yellin ~\cite{Yellin:2002xd} is used to set a limit on the DM cross section
for unknown background. With its low nucleus recoil threshold of 30.1eV, the CRESST-III detector is sensitive to DM masses larger than 188 MeV assuming the standard parameters for the DM velocity distribution, Eq.~\ref{velo_std}. Moreover, DM masses as low as 160 MeV can be probed when taking into account energy resolution. 

\begin{figure}[htb]
\centering
\includegraphics[scale=0.65]{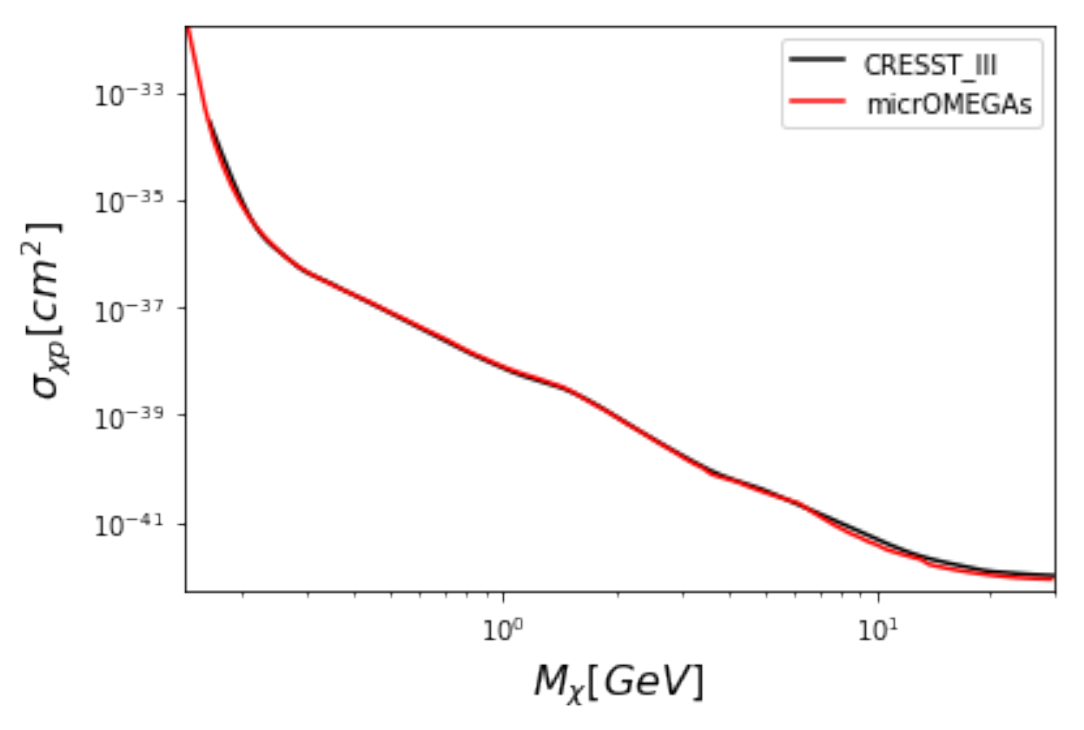}
\vspace{-.3cm}
\caption{ Comparison of the reconstructed exclusion for SI interactions from micOMEGAs with CRESST-III. }
\label{CRESST-reconstruction}
\end{figure}

Using the Optimum Interval method ~\cite{Yellin:2008da} and the  data presented in ~\cite{Abdelhameed:2019mac},  we
have recasted the exclusion limit of CRESST-III.
We use the total exposure of 5.594 kg$\cdot$days and take into account the cut-survival probability and  the acceptance  for each nucleus 
shown in Fig. 6 and  Fig.4 of ~\cite{Abdelhameed:2019mac}. The energy resolution  was considered as a free parameter which was fitted to
get the best agreement with CRESST-III  low masses exclusion. Namely  we use  a Gaussian  with $\sigma=5.5$ eV
and with a cut at $2\sigma$.   In this manner the 90\% exclusion cross section of CRESST-III is reproduced
with 10\% precision , see Fig.~\ref{CRESST-reconstruction}. 

\section{Spin-dependent interactions: recasting experimental exclusions}
\label{SDinteractions}

 In general,   SI and SD interactions  on a given atom lead to  very similar  recoil energy spectra.  Their difference is typically around 5\% and is  due only to the small  momentum dependences  
of the  SI and SD nucleus form factors. Thus experimentalists use the same set of cuts and the same background estimation for both SI and SD
interactions. It is therefore justified to use the recasting done for SI interactions and  apply it directly to SD interactions. Because there is a strong dependence on the SD form factors, to perform the recasting we use the same set of form factors as each experiment. These were obtained in ~\cite{Klos:2013rwa} and ~\cite{Fitzpatrick:2012ix} and  we cite them  here as {\tt SHELL} and {\tt EFT} respectively. Moreover, for the first the authors derive the theoretical uncertainty, we also compare our results with those obtained with the minimal form factors leading to the more robust exclusion, we cite this minimal set as {\tt SHELL-min}, see Appendix ~\ref{SDmin}.
 When we derive the 90\% limit on SD cross sections using the same SD form factors used in each  experiment, we find that our limit agrees with the experimental result with the same level of accuracy  found for SI interactions  as will be described below.

\begin{figure}[htb]
\centering
\includegraphics[scale=0.65]{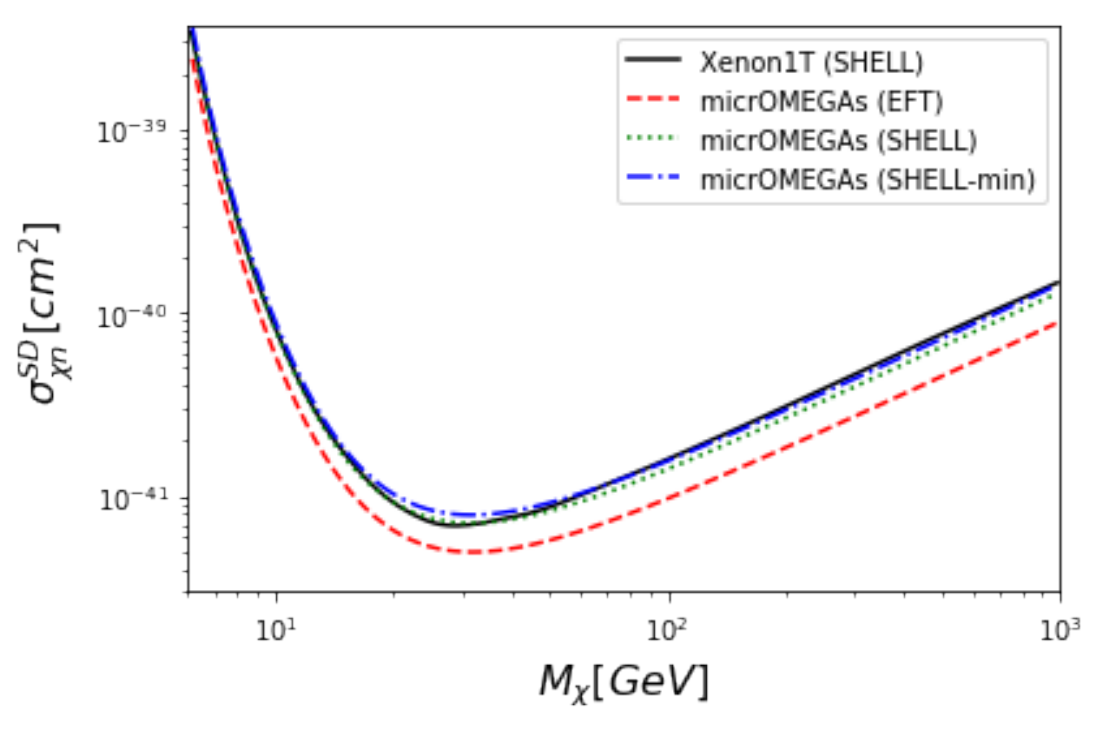}
\includegraphics[scale=0.65]{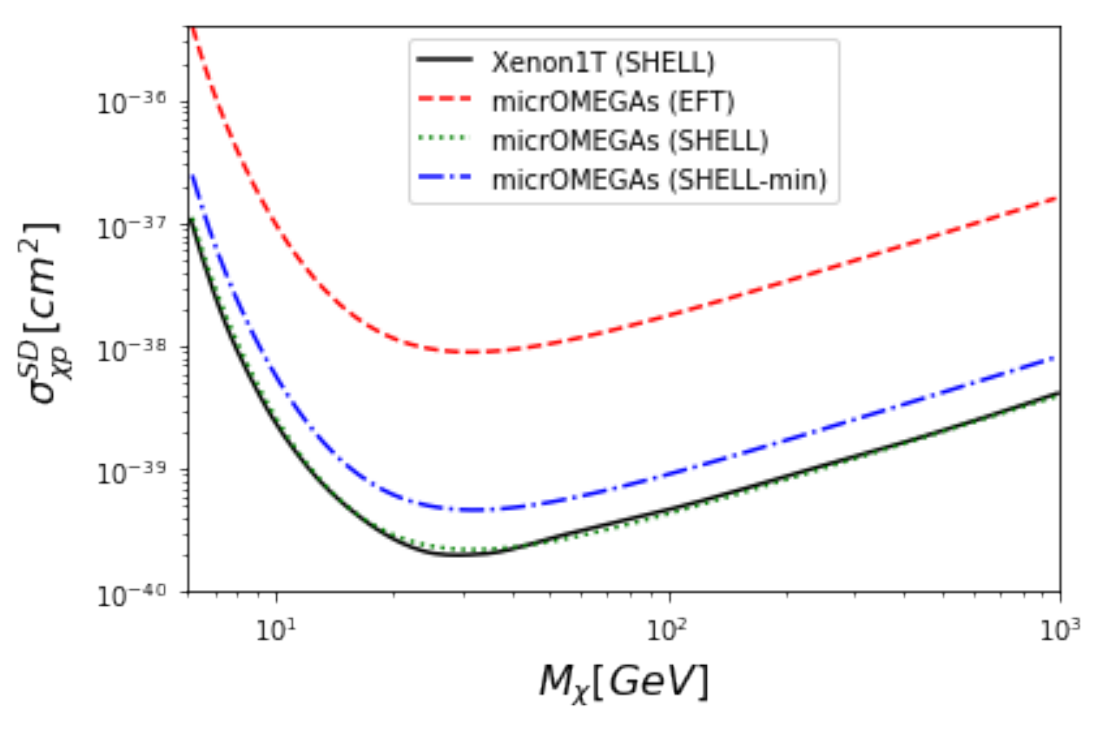}
\vspace{-.3cm}
\caption{  Comparison of the recasted 90\% limit on $\sigma^{\rm SD}_{\chi n}$ (left) and $\sigma^{SD}_{\chi p}$ (right ) from micrOMEGAs  with the XENON1T limits~\cite{Aprile:2019dbj} (black)  
with different choices  of form factors :  {\tt SHELL} (green/dot),
 {\tt SHELL-min}  (blue/dash-dot) ~\cite{Klos:2013rwa} and {\tt EFT}~\cite{Fitzpatrick:2012ix} (red/dash).}
\label{XENON1T_SD}   
\end{figure}

First we derive the 90\% limit on  SD cross sections on neutrons and protons for XENON1T, for this we take the {\tt SHELL}
 SD form factors~\cite{Klos:2013rwa}  which are also used by XENON1T. We find that an agreement  below the 15\% level with the limits on both $\sigma^{SD}_{\chi n}$ and  $\sigma^{SD}_{\chi p}$, see Fig.~\ref{XENON1T_SD}. Taking into account the uncertainty on these form factors has little impact on  $\sigma^{SD}_{\chi n}$, but weakens the limit on $\sigma^{SD}_{\chi p}$ by roughly a factor 2. The form factors {\tt EFT} lead to a more stringent limit on $\sigma^{SD}_{\chi n}$ while the limit on $\sigma^{SD}_{\chi p}$ weakens by more than one order of magnitude. Note however that XENON1T has a much lower sensitivity to $\sigma^{SD}_{\chi p}$, indeed Xenon has an even number of protons and their spin nearly cancel each other leading to small SD proton form factors.
  
\begin{figure}[htb]
\centering
\includegraphics[scale=0.65]{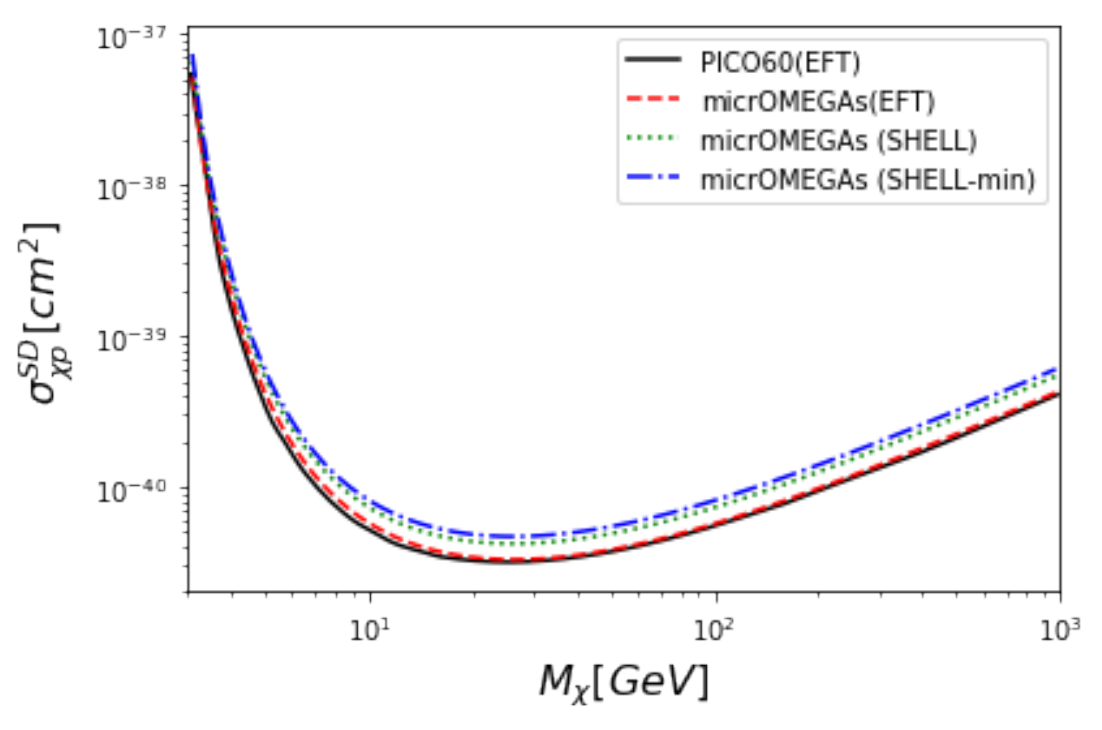}
\includegraphics[scale=0.65]{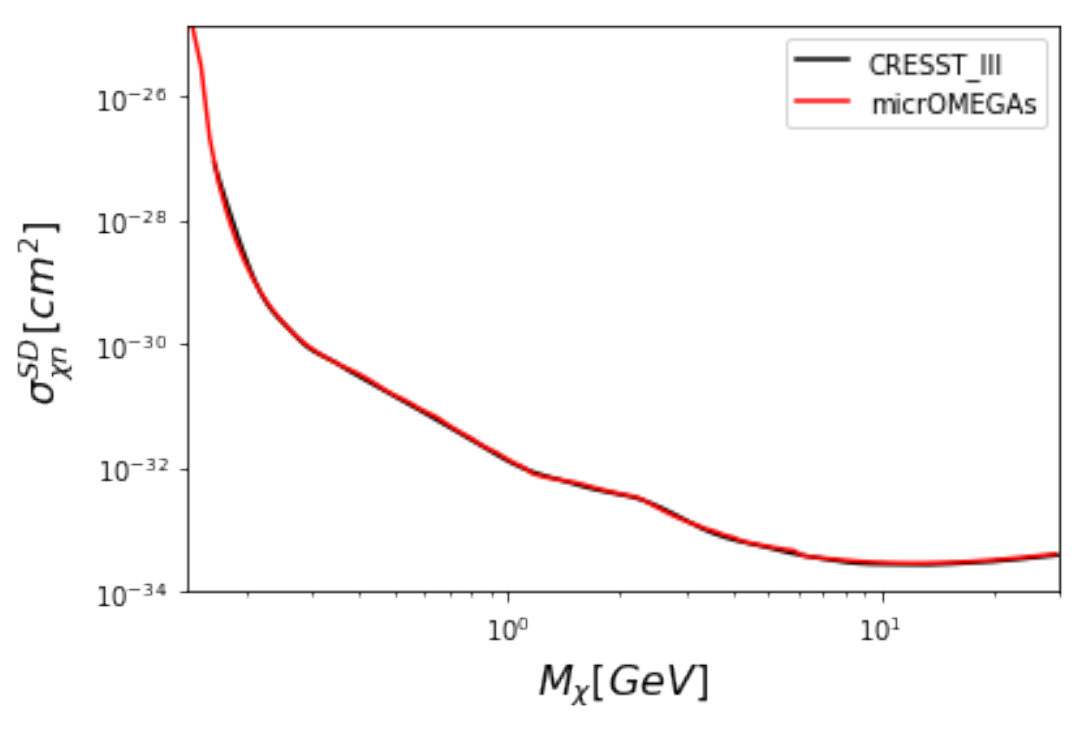}
\vspace{-.3cm}
\caption{ Left : 
        Comparison of the recasted 90\% limit on $\sigma^{\rm SD}_{\chi p}$ from micrOMEGAs (red) with the PICO-60
limit~\cite{Amole:2017dex,Amole:2019fdf} (black) using the {\tt EFT form factors} from Ref.~\cite{Fitzpatrick:2012ix}. 
         The impact of the choice of form factor is illustrated for the {\tt SHELL} (green-dot) and {\tt SHELL-min}  form factors (blue-dot-dash). Right: Comparison of the recasted 90\% limit on    $\sigma^{\rm SD}_{\chi n}$ from micrOMEGAs (red) with the CRESST-III limit~\cite{Abdelhameed:2019mac} (black) with zero momentum form factors.
         }
\label{PICO_SD}   
\end{figure}

Using the PICO acceptance described in Section~\ref{sec:pico}  we derive  the 90\% limit on SD cross section on protons and compare it with the limit presented by the PICO
collaboration~\cite{Amole:2017dex, Amole:2019fdf}, see Fig.\ref{PICO_SD}-left. For this we choose the  form factors {\tt EFT} also used by the experiment.    Our reconstruction reproduces the PICO-60 exclusion within 10\%, which is  roughly the same precision that was obtained for  SI interactions.  To check  the impact of the choice of form factors, we have also derived the exclusion using the  {\tt SHELL} and  {\tt SHELL-min}  form factors.
This weakens significantly the limit at low DM masses, up to  a factor 2 at 4 GeV, while the effect is much more moderate for DM masses above 100 GeV. The difference with the {\tt EFT} set remains below 10\% (35\%) for the {\tt SHELL} (minimal) form factors.

CRESST-III is sensitive to spin-dependent DM-neutron interactions through the $^{17}O$ isotope despite its small abundance of 0.0367\%. For this isotope, the SD form factor is only known in the zero momentum 
limit ~\cite{Bednyakov:2006ux}, we take  the spin expectation $\langle S_n\rangle=0.5$. Following the same procedure as for SI interactions, we derive the recasted 90\% limit on $\sigma^{\rm SD}_{\chi n}$  and in  Fig.~\ref{PICO_SD} - right, we make a comparison with  the results of 
CRESST-III ~\cite{Abdelhameed:2019mac}, the discrepancy is below $10\%$. \footnote{Note that the preliminary results for the SD exclusion
~\cite{Abdelhameed:2019hmk}  were
improved in ~\cite{Abdelhameed:2019mac}.} The agreement with CRESST-III for the exclusion is at the same level  as for the SI case.

\section{Applications}
\label{Applications}
In this section we show how to exploit our reconstruction of DD experimental limits to obtain limits on specific  DM models  while taking into account uncertainties from astrophysical and nuclear physics parameters.  All numerical results presented below can be easily reproduced  with the  micrOMEGAs code.
The corresponding code is stored in \verb|mdlIndep/dd_exp.c| of micrOMEGAs.

\subsection{The case of a light mediator.}

\begin{figure}[htb]
\centering
\includegraphics[scale=0.65]{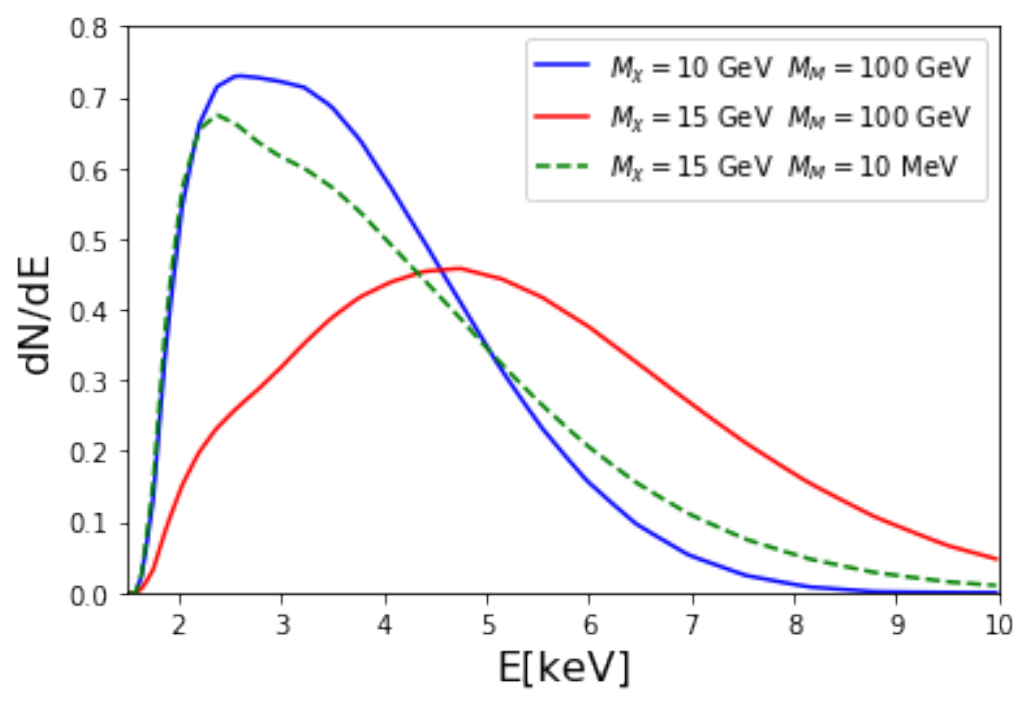}
\includegraphics[scale=0.65]{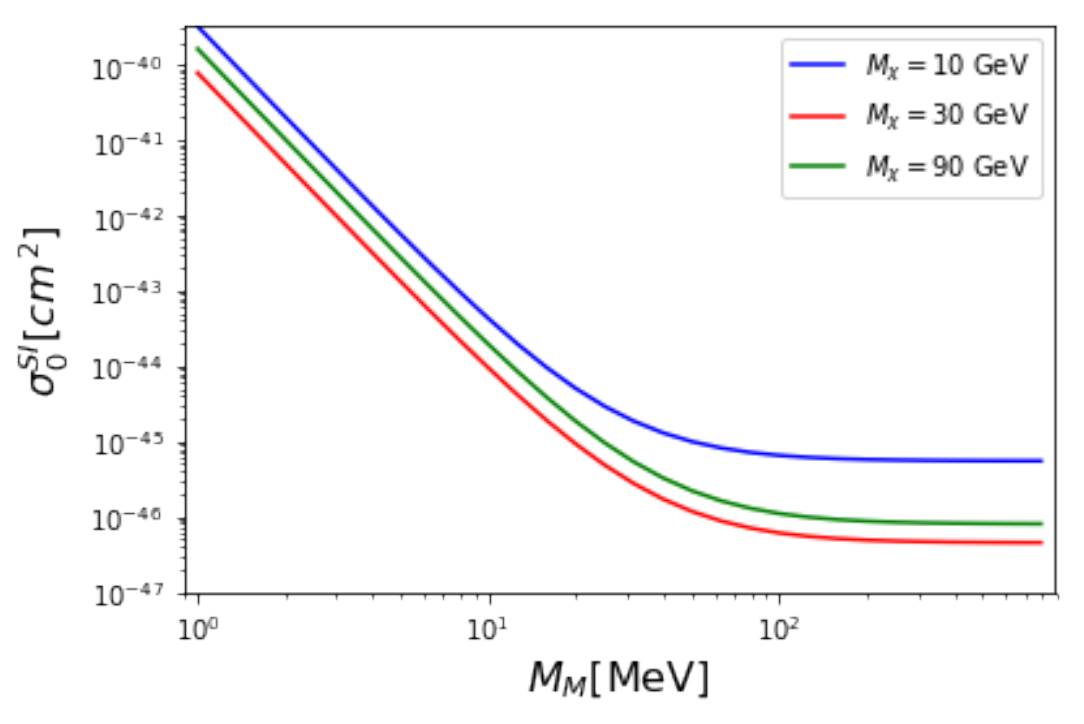}
\vspace{-.3cm}
\caption{Left : Comparison of the  nucleus recoil distributions after folding in the XENON1T acceptance, $p_{eff}$,  for a heavy mediator  $M_M=100$~GeV (full) and  $M_{\chi}=10,15$~GeV with that of a light mediator, $M_M=10$~MeV (dash)  and $M_{\chi}=15$~GeV.   Right:  Dependence of the excluded cross section on the mediator mass for $M_\chi=10,30,90$~GeV.
}
\label{mediatorNR}
\end{figure}

When  DM-nucleus interactions are due to the exchange of a light mediator in   t-channel,
the  standard formula that relates the DM-nucleon cross section at zero momentum with the recoil energy distribution cannot be applied.  
Indeed it rests on the assumption that the mass of the mediator is much larger than the Mandelstam variable
 $t=- 2 M_A E_R$ where $E_R$ is the nucleus recoil energy and $M_A$ the mass of the recoiling nucleus. For the typical minimal recoil energy $E_R\approx
2 {\rm keV}$ and $M_{Xe}$=130GeV this corresponds to $t=-(22{\rm MeV})^2$.  Thus for mediator masses significantly below 1 GeV, 
an additional factor describing  the t-dependence should be included. The recoil energy distribution from DM-nucleus elastic scattering is then replaced with 
\begin{equation}
\label{LowMassMed}
\frac{dN_{A}^{light}}{dE} = \frac{M_M^4}{(M_M^2 +2 M_{A} E)^2} \frac{dN_{A}^{std}(\sigma_0)}{dE} 
\end{equation}
where $N_{A}^{std}$ is the standard expression for the number of recoil events for a point-like
interaction, Eq.~\ref{eq:SI},~\cite{Lewin:1995rx} with elastic scattering cross section $\sigma_0$, $M_M$ is the mass of the t-channel mediator.  Taking into account  the  contribution of the  transfer momentum in the  propagator of the  light mediator leads to an overall decrease of the recoil signal
and to a shift towards lower energies. This can be seen in  Fig.~\ref{mediatorNR} (left) where the signals for a  
DM with mass of 15 GeV are compared in the case of  a light mediator $M_M=10~{\rm MeV}$ and a heavy mediator,  $M_M=100~{\rm GeV}$.  Moreover the recoil spectrum with the light mediator is shown to be very similar to the one for $M_\chi=10$~GeV and $M_M=100~{\rm GeV}$. These signals include the reconstructed acceptance of XENON1T, $p_{eff}$, and are obtained for $\sigma^{90}$, to ease the comparison the distribution for the light mediator includes a normalisation factor. 

\begin{figure}[htb]
\centering
\includegraphics[scale=0.65]{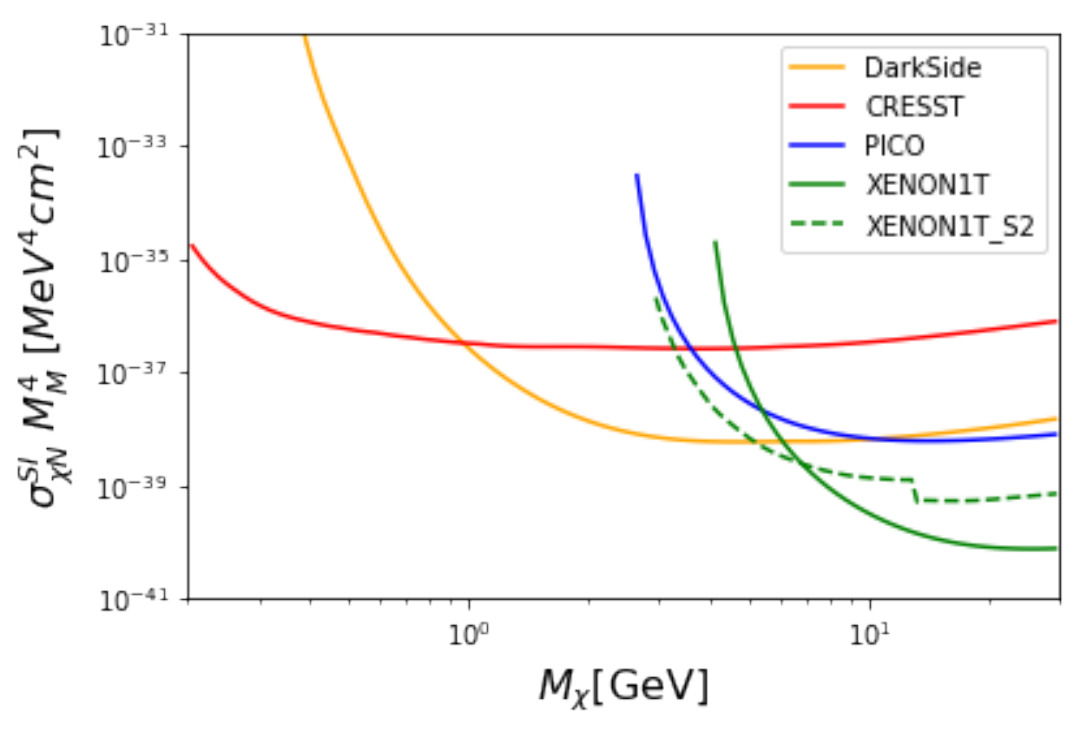}
\vspace{-.3cm}
\caption{Limits on the spin-independent DM nucleon point-like cross section  for a light mediator,  $M_M=10$~MeV, using the micrOMEGAs recast of XENON1T, DarkSide-50, PICO-60 and CRESST-III. The limit derived by XENON1T using a ionization-only analysis is also displayed, XENON1T-S2.~\cite{Aprile:2019xxb}.
}
\label{fig:light_mediator}
\end{figure}

In any model with a light mediator, we can use Eq.~\ref{LowMassMed} to calculate the recoil energy signal and extract the
 dependence of the 90\% excluded cross section  on the mediator mass.
 The zero velocity  excluded cross section ($\sigma_0$) is displayed  in  Fig.~\ref{mediatorNR} (right) for XENON1T and for different  DM masses.    As  expected,  the mediator mass dependence comes into play at $M_M=100$MeV and the effect is significant at 50MeV.  For very small mediator masses, all DM masses have a similar dependence on $M_M$, the reason is that the key ingredient in setting the limit is the detector threshold.  The model independent limits on SI interactions  in the case of a light mediator obtained from the micrOMEGAs recasting are compared in Fig.~\ref{fig:light_mediator} for different experiments. Moreover the limit derived by the XENON1T collaboration using a S2 only analysis that allows to extend the sensitivity to lower masses is also shown for comparison~\cite{Aprile:2019xxb}.

\begin{figure}[htb]
\centering
\includegraphics[scale=0.65]{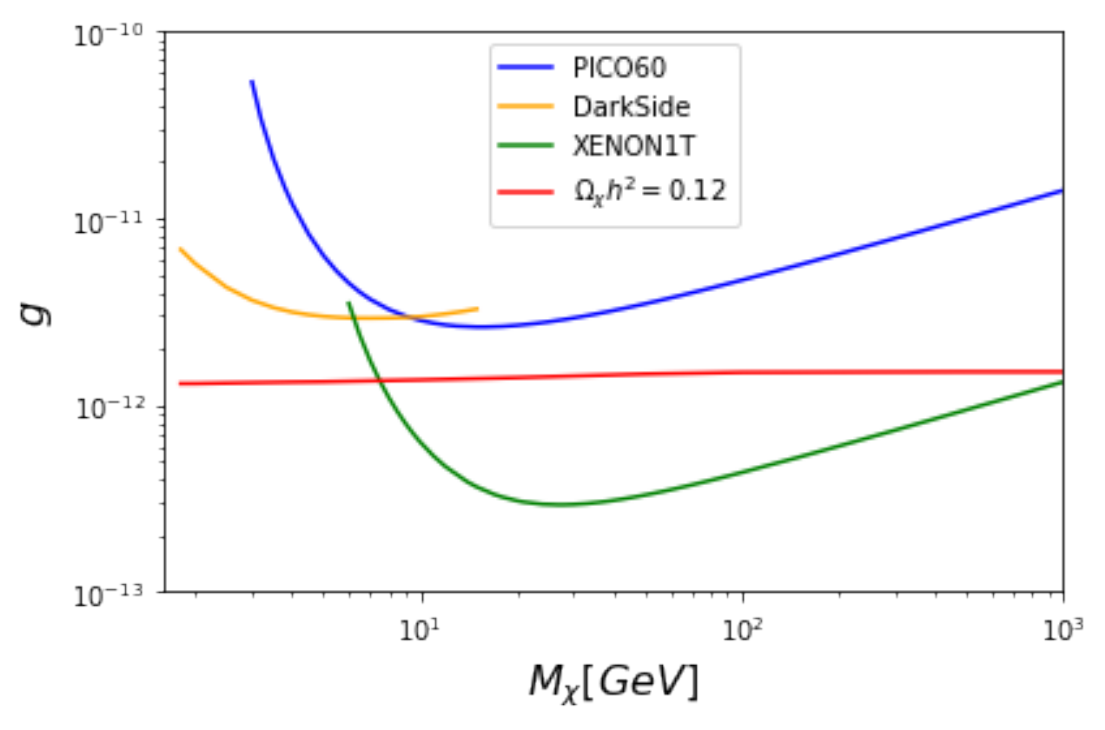}
\includegraphics[scale=0.65]{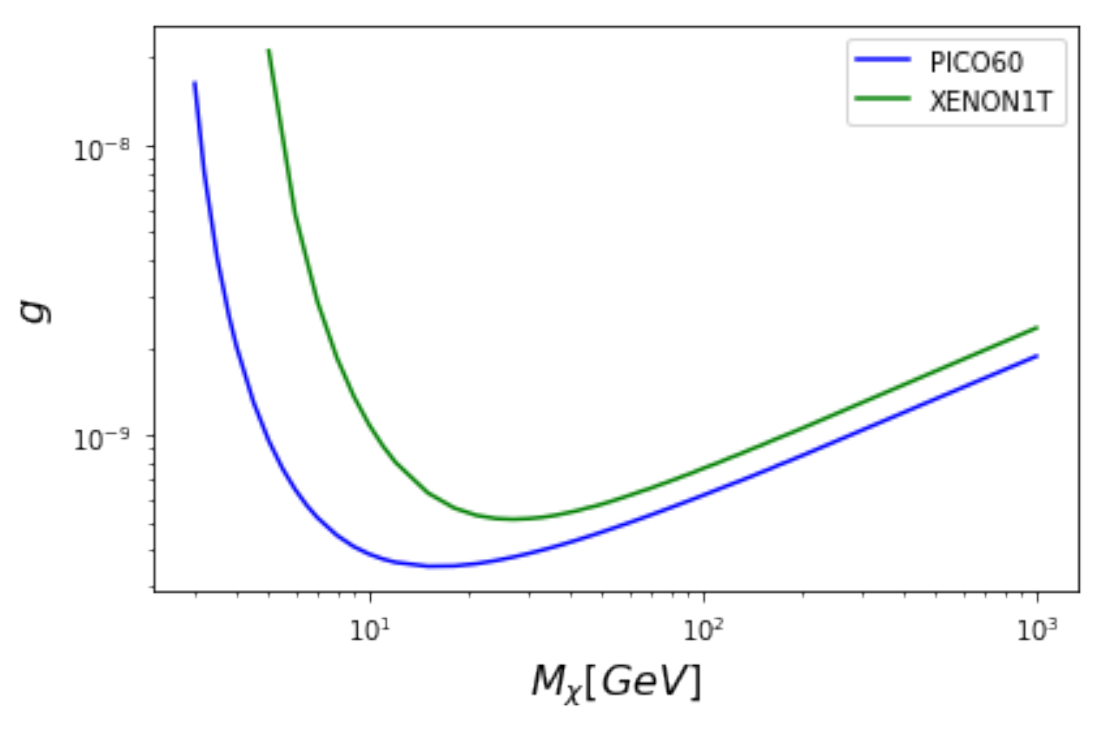}
\vspace{-.3cm}
\caption{ Limits on the Z' coupling, $g=g_f g_\chi$,  from DarkSide-50, PICO-60, and XENON1T, for the Z' model with pure vector couplings (left) and from PICO-60 and XENON1T for pure axial-vector couplings (right)  for a light mediator, $M_Z'= 1$~MeV. }
\label{fig:Zprime_limits_SI}
\end{figure}

To illustrate the effect of the light mediator on the direct detection exclusion in a specific model we consider the case of a Z' mediator with a universal coupling to SM fermions,  
\begin{equation}
{\cal L} = - Z'_\mu \left(g_\chi \bar\chi\gamma^\mu\chi   +  g'_\chi   \bar\chi\gamma^\mu\gamma^5 \chi \right)  - \sum_f   Z'_\mu \left( g_f  \bar f \gamma^\mu f +g'_f \bar f \gamma^\mu\gamma_5 f  \right)
\end{equation}
We assume either pure vector couplings ($g'_\chi=g'_f=0$)  or axial-vector  ($g_\chi=g_f=0$) couplings which give rise respectively to SI and SD interactions. We further assume identical couplings to all fermions $f$. 
The results are displayed in  Fig.~\ref{fig:Zprime_limits_SI} for both SI interactions and SD interactions. For SI interactions, the region that is compatible with the measured value of the relic density is excluded by XENON1T for $M_\chi> 8$~GeV, this region corresponds to $g=g_\chi g_f \approx 1.4 \times 10^{-12}$ with less than 10\% variation over the mass range considered.  For SD interactions, the current experiments cannot yet probe the preferred value for the relic density, as the couplings probed are roughly three orders of magnitude larger than the ones required by the relic density ($g \approx 1.4 \times 10^{-12}$ ). Moreover CRESST-III only probes values of couplings $g > 10^{-7}$, thus the corresponding limits are not displayed. Note that  Fig.~\ref{fig:Zprime_limits_SI} shows the best limit whether it comes from SD interactions with protons (PICO) or neutrons (XENON1T). 

\subsection{Millicharged Dark matter}

Millicharged DM which interacts with the SM through photons provides another example of a light mediator, the massless photon in this case.  Typically  a kinetic mixing between a new gauge boson and the hypercharge leads to DM interacting with the photon with a millicharge, $q_{\chi}$, ~\cite{Haas:2014dda}
\begin{equation}
{\cal L}= q_{\chi} e \bar\chi\gamma^\mu \chi A_\mu 
\end{equation}
where we have omitted the terms that describe interactions with the new gauge boson. The recoil energy distribution  for DM nucleus elastic scattering is similar to the one for the light mediator, Eq.~\ref{LowMassMed}, 
\begin{equation}
\label{eq:millicharge}
\frac{dN_{A}^{m}}{dE} = \frac{M_{ph}^4}{(2 M_{A} E)^2} \frac{dN_{A}^{std}(\sigma_0)}{dE} 
\end{equation}
where
\begin{equation}
\sigma_0= 16 \pi \alpha^2_{EM} q^2_\chi Z_A^2 \frac{\mu^2_{\chi p}}{M^4_{ph}}
\label{eq:mph}
\end{equation}
and $M_{ph}$ is a  parameter  with mass   dimension  which does not enter the final result.

\begin{figure}[htb]
\centering
\includegraphics[scale=0.65]{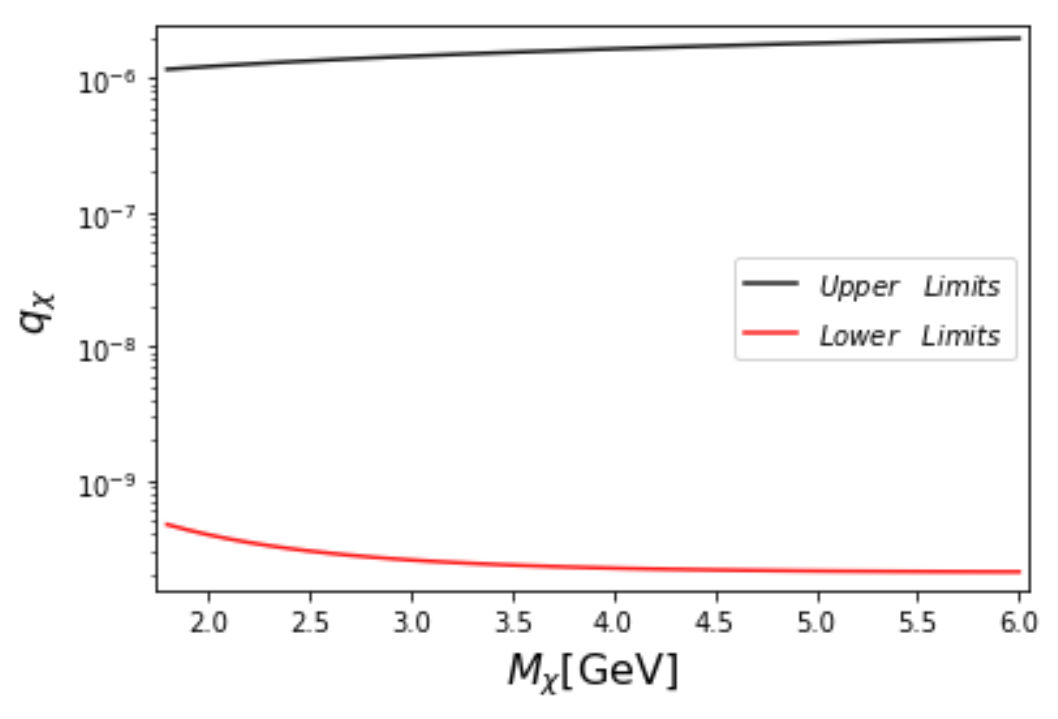}
\includegraphics[scale=0.65]{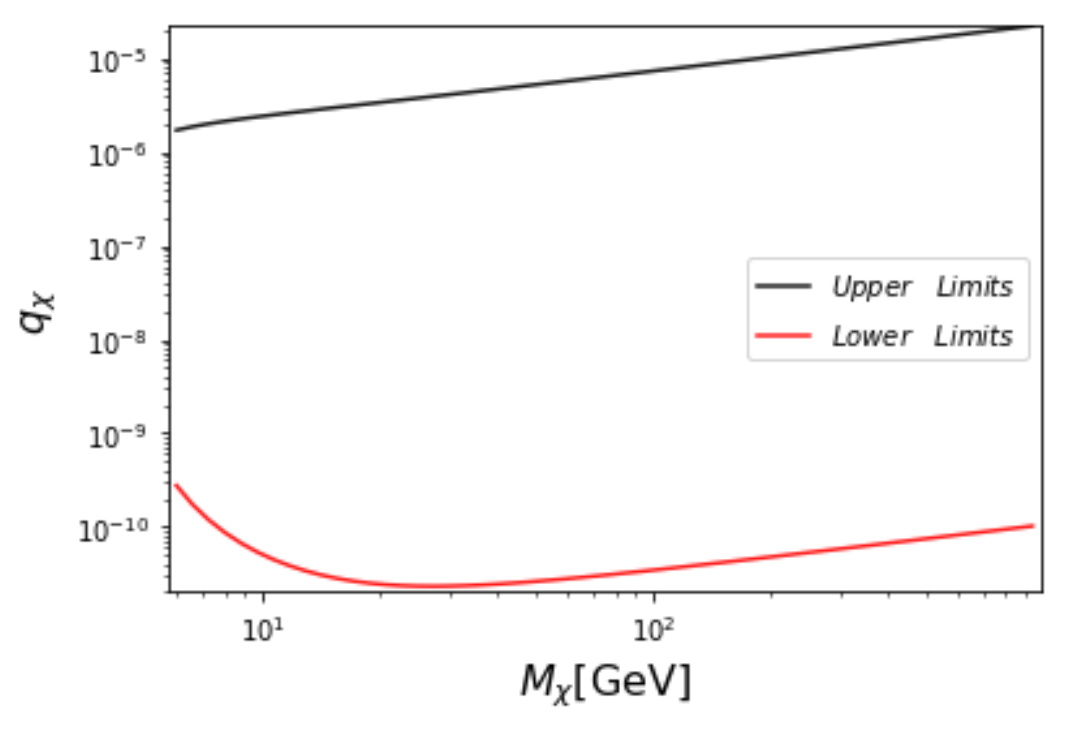}
\vspace{-.3cm}
\caption{ The 90\% exclusion on the   DM millicharge $q_{\chi}$ as a function of the DM mass 
using recasted results of DarkSide-50 (left) and of  XENON1T (right). The region above the top curve cannot be probed by underground DD experiments. }
\label{millichargeFig}
\end{figure}

The 90\%  lower  limits on  $q_{\chi}$   obtained after imposing the  DarkSide-50 and  XENON1T limits  are   presented in  Fig.\ref{millichargeFig}. 

 Direct detection  experiments cannot probe large values of the charge, $q_\chi$, since a  millicharged DM will  loose energy through its interaction with rocks before it reaches the detector.  
Elastic scattering of DM particles with atomic nuclei is the main process responsible for energy loss. 
The cross section for elastic scattering reads \cite{Landau1981Quantum}
\begin{equation}
\label{dSigmadCos} 
  \frac{d\sigma}{d\cos{\theta_{cm}}}=2\pi\left| \frac{2\mu}{q} \int \limits_0^\infty V(r)r
\sin(rq)dr\right|^2
\end{equation}
where $\mu$ is the reduced mass of colliding particles and $q$ is the transfer momentum. For a nucleus charge  screened by electrons, the potential is given by 
\begin{equation}
   V(r)=\frac{q_\chi  Z_A e^2}{r}e^{-r/R_A}
\end{equation}
where the atomic radius is approximated by $R_A \approx 0.8853 Z^{\frac{1}{3}}_A/m_e \alpha_{em}$. Note that this rough approximation is sufficient since 
the energy loss   depends only logaritmically on    $R_A$.
The energy loss, $E_{loss}$ of one millicharged particle in an elastic collision with a nucleus is obtained after 
integrating Eq.\ref{dSigmadCos}, 
\begin{equation}
\label{EsigmaLost}
  \langle E_{lost}\sigma_A \rangle =2\pi \frac{(q_\chi Z_A e^2)^2}{v^2 M_A} \left( \log(1+(2v\mu_{\chi A} R_A)^2) -\frac{(2v\mu_{\chi A} R_A)^2}{1+(2v\mu_{\chi A} R_A)^2}\right)
\end{equation}
The energy, $E_\chi$ of a DM particle passing through the Earth  is then given by
\begin{equation}
  \frac{dE_\chi}{dx} = -\sum_{A \in Earth} \langle E_{lost}\sigma_A \rangle n_A
\end{equation}
where $n_A$ is the  number density of the element A in the  Earth and $x$ is the distance from the surface.
 If the DM mass is above the  GeV scale then $2v\mu_{\chi A} R_A \gg 1$ and $ \frac{dE_\chi}{dx}\approx \frac{C}{E_\chi}$.  Thus,  at some finite distance from the Earth surface, the DM will stop and  drift towards  the Center of the Earth driven by gravitational interactions. For DM to be detectable its energy must be above the 
  detector threshold $E_{tr}$, thus the condition that a DM with maximum velocity $v=v_{esc}+v_{Earth}$ will reach the detector located at a distance $H$ below the surface of the Earth with $E_\chi>E_{tr}$ leads to a linear equation in $q_\chi^2$
\begin{equation}
    \label{kinCond}
     \int \limits^{E_{min}}_{E_{max}}  \frac{dE_\chi}{{dE_\chi}/dx} = H  \;\; {\rm where} \;\;
      E_{min} =  \frac{E_{tr}M_A M_{\chi}}{4 \mu_{\chi A}^2} \;\; ;\;\;
      E_{max} = \frac{M_{\chi}}{2}(v_{Esc}+v_{Earth})^2 
\end{equation}  
The corresponding  upper limit on the millicharge excluded by either DarkSide-50 or XENON1T is at least three orders of magnitude above the respective lower limits, see Fig.~\ref{millichargeFig}. Here  we used
H=1400 m and  $E_{tr}= 0.1(1.6)$ keV for DarkSide-50 (XENON1T) .

\subsection{Dependence on DM  distributions}
\label{DMvelocity}

As mentioned previously, most experiments publish their results assuming that the DM velocity distribution in the neighbourhood of the Sun  is a Maxwell distribution with parameters
given in Eq.\ref{velo_std}. However,   the recent estimates for $\rho_{\chi}$ point to a slightly larger value~\cite{Tanabashi:2018oca, Salucci:2010qr}
\begin{equation}
    \rho_{\chi}=  (0.39 \pm 0.03)(1.2 \pm 0.2)(1\pm \delta_{triax}){\rm GeV/cm^3}
\end{equation}
where $\delta_{triax}< 0.2$. Clearly, since   $\rho_{\chi}$ is just an overall factor,  changing its value will amount to simply  rescaling  the 90\% excluded cross section by a factor  of $\rho_{\chi}/0.3$.
 
 To estimate the impact of the DD limits on the parameters of the velocity distribution we have varied the parameters of the Maxwell distribution within the  range~\cite{Green:2017odb,Monari_2018,Wu:2019nhd}
\begin{equation}
\label{velocity_intervals}
  v_{Rot}=220\pm 18  \frac{km}{s}\;\;\;\;v_{Earth}=232-252 \frac{km}{s}\;\;\;\;v_{Esc}=580\pm 63 
\frac{km}{s}  \;\;\;\;  \rho_\chi=0.468\pm 0.202
\end{equation}

\begin{figure}[htb]
\centering
\includegraphics[scale=0.65]{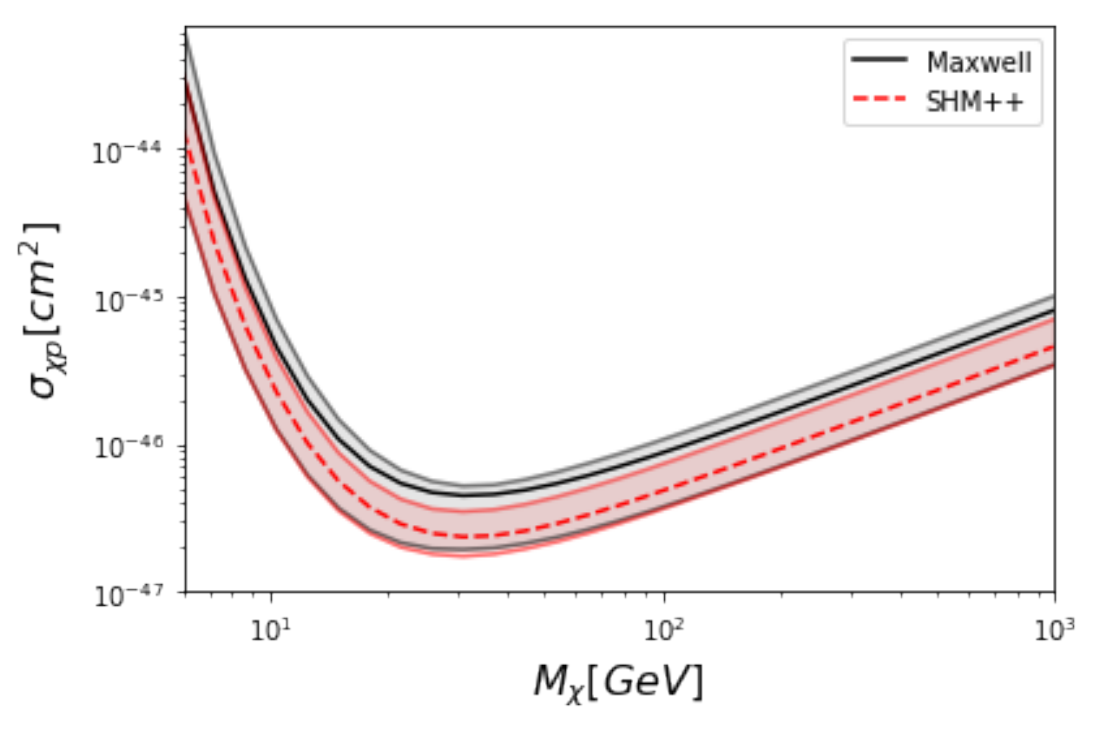}
\includegraphics[scale=0.65]{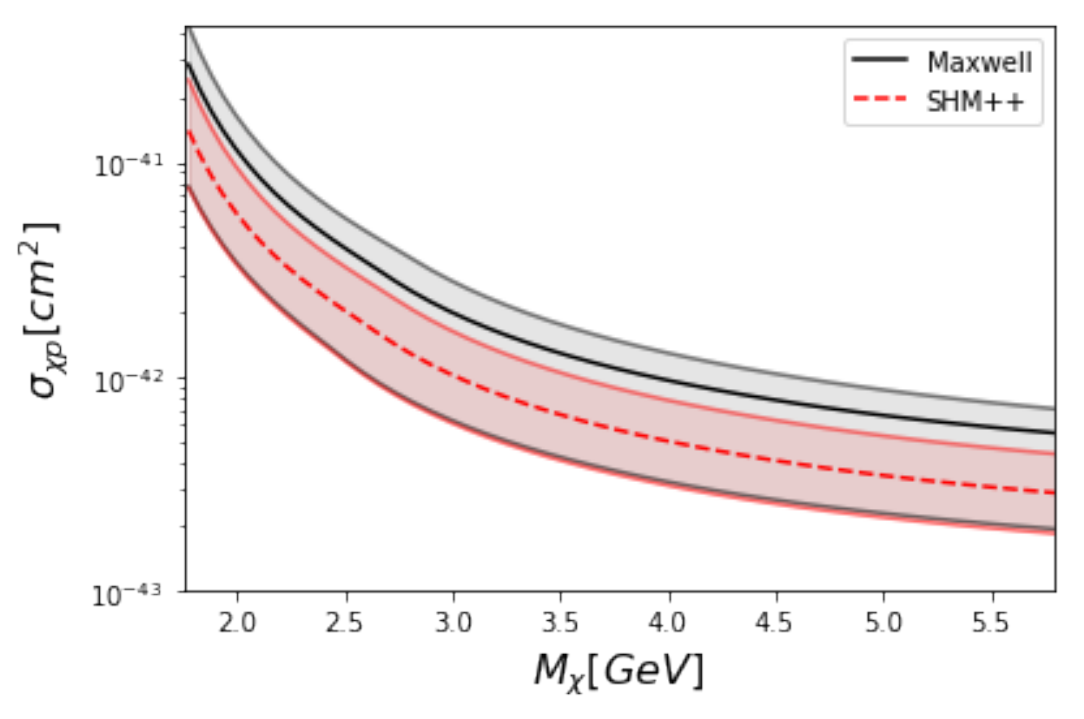}
\vspace{-.3cm}
\caption{ Influence of the uncertainty on the DM velocity  distribution on XENON1T (left) and DarkSide-50 (right) 90\% excluded SI cross sections. The allowed region for the Maxwell distribution (shaded black) and the SHM{\scriptsize{++}} distribution (shaded red) together with the full curves showing the standard central values for both distributions. 
  }
\label{velo}
\end{figure}

\begin{figure}[hbt]
\centering
\includegraphics[scale=0.65]{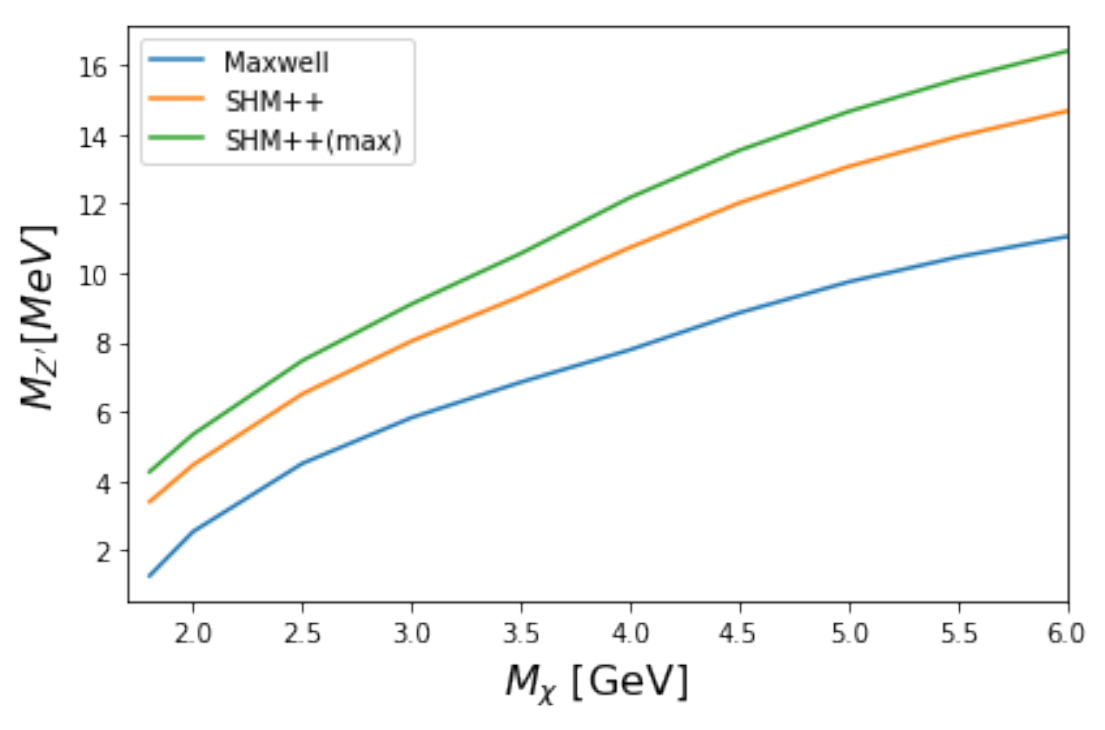}
\includegraphics[scale=0.65]{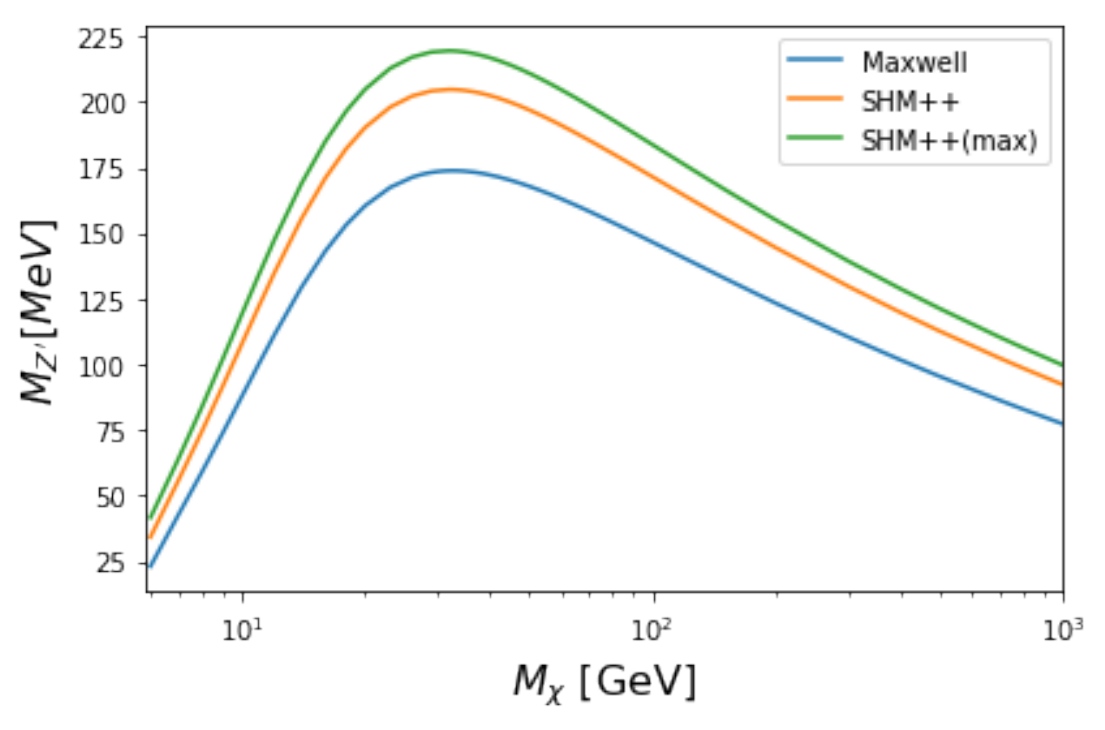}
\vspace{-0.3cm}
\caption{Impact of the velocity distribution on exclusion for  the Z' model
with vector couplings from DarkSide-50 (left) and XENON1T (right) in the $M_{Z'}-M_\chi$ plane.  Here $g_{Z'}=1\times 10^{-7}$, $g_\chi=5.5\times 10^{-5}$. The exclusion obtained with the Maxwell distribution and standard parameters is compared with the ones  obtained with SHM{\scriptsize{++}} with standard values for the parameters and with the parameters leading to the most severe constraint (SHM{\scriptsize{++}}(max)). The regions below the curves are excluded.
}
\label{veloZp}
\end{figure}

The strongest and weakest 90\% excluded cross sections for the XENON1T experiment for these  intervals  are shown  in Fig.~\ref{velo} together with the exclusion corresponding to the standard parameters in Eq.~\ref{velo_std}.  For DM masses above roughly 10 GeV, most of the variations in the exclusion limit is due to  $\rho_{\chi}$, in particular  the upper $1\sigma$ range leads to a more aggressive limit by about a factor 2 while the limit is weakened by around 10\%  when using the lowest value for  $\rho_{\chi}$. Roughly another 10\% shift in the limit is due to the variation of other parameters. For low DM masses, corrections can be much larger. 
For example for $M_{\chi}\approx 6~{\rm GeV}$ the excluded cross section increases  by more than a factor 2. This is mainly due to a decrease in  $v_{Esc}$  which requires a heavier DM to pass the threshold for nuclear recoils. For the same reason an increase in  $v_{Esc}$ leads to a more aggressive limit. 
An alternative DM distribution which is compatible with {\it Gaia} data was suggested in Ref.~\cite{Evans:2018bqy},  
it leads to more stringent limits at all masses since the main difference with the Maxwell distribution  is in the much larger  central value for $\rho_{\chi}$. 
We have also varied the parameters of the SHM{\scriptsize{++}} distribution within their $1\sigma$ range defined in 
Eq.~\ref{eq:SHMpp} and found a near overlap of the most stringent exclusion with that of the Maxwell distribution.  Again, $\rho_\chi$ and $v_{esc}$ are the parameters that have the largest impact on the exclusion limit. After factoring out the linear dependence on $\rho_\chi$, we still find that the limit shifts by more than a factor 2 for $M_{\chi}\approx 7~{\rm GeV}$ and by about 20\% for $M_{\chi}> 200~{\rm GeV}$.
Similar conclusions are obtained for the DarkSide-50 exclusions in the low mass region, see
Fig.~\ref{veloZp}- right.

We also examine the impact of the velocity distribution on the exclusion limit for the simplified Z' model with vector couplings introduced in the previous section. 
Fixing the values of the couplings to $g_{Z'}=1\times 10^{-7}$ and $g_\chi=5.5\times 10^{-5}$ we show how much the exclusion
 on the Z' mass from DarkSide-50 and XENON1T can be reinforced assuming an aggressive exclusion with the SHM{\scriptsize{++}} distribution. The latter, labelled SHM{\scriptsize{++}}(max) corresponds to the upper value of the $1\sigma$ range  for the parameters $\rho_\chi,{\mbox{vrot}}, {\mbox{vesc}}$ in Eq.~\ref{eq:SHMpp}. With this choice the lowest limit on $M_{Z'}$ increases  by more than a factor 2 for $M_\chi=1.8$~GeV to about 40\% for 
$M_{\chi}> 100$~GeV as compared to the Maxwell distribution with standard parameters, Eq.~\ref{velo_std}. This confirms our expectations that the impact of the velocity distribution is more important for spectra peaked at low energies.

\section{Conclusion}

 In this paper we demonstrate how the results from recent DM direct detection experiments can be applied to  DM models with features that can somewhat differ from the ones assumed when deriving the experimental limits. 
After validating the recast of experimental exclusions, we illustrated how these can be applied to specific DM models, in particular models with a light mediator or a millicharged DM for which the spectrum of nuclear energy recoil is shifted towards low energies from the one of a heavy mediator. We also illustrated the impact of the choice of nuclear form factor for spin dependent interactions and of the choice of velocity distributions. These recasts can also be used to derive direct detection limits on multicomponent DM. 
 These recasts are available in micrOMEGAs  which  contains  new routines that provide the exclusion cross section for the direct detection experiments that provide the best exclusion for spin independent and spin dependent interactions for DM masses from 160~MeV upto the TeV range. 
Note that these recasts  can be used with any of the generic models implemented in micrOMEGAs and that as for all direct detection routines apply to models where the direct detection cross section can be described by the  low-energy Lagrangians for fermion, scalar or vector DM  listed in Ref.~\cite{Belanger:2008sj} extended to the case of light mediators. These routines will be extended to include future experimental limits as they become available.

\section{Acknowledgements}

We have benefited from exchanges with members of various direct detection collaborations, in particular Masayuki Wada and Davide Franco (DarkSide), Victor Zacek and Scott Fallows (PICO) and   Florian Reindl (CRESST).
We also acknowledge useful discussions with Bryan Zald\'ivar. 
This work  was funded by RFBR and CNRS, project number 20-52-15005. The work of A. Pukhov was supported in part by a grant AAP-USMB and by 
 the Interdisciplinary Scientific and Educational School of Moscow University "Fundamental and Applied Space Research" .
The authors would like to thank the Mainz Institute for Theoretical Physics, the ICTP-SAIFR in Sao Paulo, and the 
Paris-Saclay Particle Symposium 2019 with the support of the P2I and SPU research departments and of
the P2IO Laboratory of Excellence (program "Investissements d'avenir"
ANR-11-IDEX-0003-01 Paris-Saclay and ANR-10-LABX-0038) for their hospitality and support during the completion of this work.

\appendix
\section{micrOMEGAs routines}   
\label{Appendix}
We describe the micrOMEGAs  routines  that can be used to extract constraints on DM models 
based on the results of the direct detection experiments.  Examples on how to use these routines can be found  in \verb|mdlIndep/dd_exp.c| of micrOMEGAs. All results presented here can also be reproduced with this code.

\subsection{Experimental data}

The SI 90\%  DD  limits  tabulated from the results presented by  XENON1T~\cite{Aprile:2018dbl}, DarkSide-50~\cite{Agnes:2018ves}, PICO-60~\cite{Amole:2019fdf}  and CRESST-III~\cite{Abdelhameed:2019hmk}  are accessible through  the following functions \\

\noindent
$\bullet$ \verb|XENON1T_90(Mdm)          | for  $ 6 <  M_{DM} < 1000$~GeV, \cite{Aprile:2018dbl}  \\
\noindent
$\bullet$ \verb|DS50_90(Mdm)             | for  $ 0.7 < M_{DM} < 15$~GeV, \cite{Agnes:2018ves}   \\
\noindent
$\bullet$ \verb|PICO60_90(Mdm)           | for  $ 3 < M_{DM} < 10000$~GeV, \cite{Amole:2019fdf}   \\
\noindent
$\bullet$ \verb|CRESST_III_90(Mdm)       |  for $ 0.35 < M_{DM} < 12$~GeV. \cite{Abdelhameed:2019hmk} \\

The corresponding SD 90\% exclusion limits are contained in the functions\\

\noindent 
$\bullet$ \verb|PICO60_SDp_90(Mdm)       | for  $ 3 < M_{DM} < 10000$~GeV,  \cite{Amole:2019fdf} \\
\noindent
$\bullet$ \verb|XENON1T_SDp_90(Mdm)      | for  $ 6 <  M_{DM} < 1000$~GeV, \cite{Aprile:2019dbj} \\
\noindent
$\bullet$ \verb|XENON1T_SDn_90(Mdm)      | for  $ 6 <  M_{DM} < 1000$~GeV, \cite{Aprile:2019dbj} \\
\noindent
$\bullet$ \verb|CRESST_III_SDn_90(Mdm)   |  for $ 0.35 < M_{DM} < 12$~GeV. ~\cite{Abdelhameed:2019hmk}\\

These functions give the  excluded cross sections in ${\rm cm}^2$ . For a DM mass outside the range specified the function returns NaN.

\subsection{Recasting the experimental limits with  micrOMEGAs}  

\noindent
$\bullet$ \verb|DD_pvalCS|(\verb|expCode|, $f_v$, $\sigma_{SI_P},\sigma_{SI_N},\sigma_{SD_P}, \sigma_{SD_N}$,\&expName)\\
calculates the  value $\alpha= 1-C.L.$  for a model with DM-nucleon cross sections 
$\sigma_{SI_P},\sigma_{SI_N},\sigma_{SD_P}, \sigma_{SD_N}$. Cross sections are specified in
[pb] units. The return value 0.1 corresponds to a 90\% exclusion.   
 The {\tt expCode} parameter can be any of the codes  \verb|XENON1T_2018,DarkSide_2018|, \verb|CRESST_2019,PICO_2019| or their combination concatenated with the symbol
$\mid$. There is also a predefined parameter that  currently combines these experiments 
\begin{verbatim}
  AllDDexp=XENON1T_2018|DarkSide_2018|PICO_2019|CRESST_2019;
\end{verbatim}

The parameter {\tt char* expName}  is used to indicate the   experiment that  provides
the best exclusion among those specified in  {\tt expCode}.   The function {\tt DD\_pvalCS } calculates the
exclusion for each experiment  independently, returns the smallest $\alpha$, and assigns the 
name of the corresponding experiment to  {\tt expName}  if it is not {\tt NULL}.  

The $f_v$ parameter specifies the DM velocity distribution in the detector frame. For
example, one can use   {\tt Maxwell} or {\tt SHMpp} which are
included in \micro~, 
\ref{veloAppendix},  otherwise the user can define another distribution.   
The DM velocity distribution  has to be normalized  as in Eq.\ref{normf}.
 The units  are $km/s$ for v and $s/km$ for  $f_v(v)$. {\tt DD\_pvalCS} implicitly depends on the 
global parameters  {\tt Mcdm}  and  {\tt rhoDM} which specify the DM mass and DM  local density respectively.

For  XENON1T one can chose  between  $p_{eff}^{q}$ with $q=0,1,2$, 
see Section \ref{Xenon1Trecast}. The  flag {\tt Xe1TnEvents=q} allows to choose the  corresponding recasting,  otherwise and by default the code uses  $p_{eff}^1$. For PICO-60, the user can choose between the recasting based on Feldman-Cousins statistics, {\tt PICO60Flag=0} 
which is the default value, or the one based on  Neyman one side belt exclusion, {\tt PICO60Flag=1}.\\

\noindent
$\bullet$ \verb|DD_factorCS|(\verb|expCode|, $\alpha$, $f_v$, $\sigma_{SI_P},\sigma_{SI_N},\sigma_{SD_P},\sigma_{SD_N}$,\&expName)\\ 
returns the overall factor which should be applied to the cross sections, $\sigma_{SI_P},\sigma_{SI_N},\sigma_{SD_P}, \sigma_{SD_N}$ to reach  the exclusion level $\alpha$.  
All parameters are the same as in {\tt DD\_pvalCS} above. \\

\noindent
$\bullet$ \verb|*dNdEFact(Enr_kev, A)|\\
is the address of the function which modifies the nucleus recoil distribution  for {\tt DD\_pvalCS}
and {\tt DD\_factorCS}   to take into account a 
t-channel propagator  with  small or zero mass.  By default \verb|dNdEfact=NULL| and this 
function does not contribute to  the calculation of the direct detection cross sections. Otherwise  it is taken as an additional factor 
in the nucleus recoil distribution, see Eq.\ref{LowMassMed}.  
The parameter {\tt Enr\_kev} is the recoil energy in [keV] units, A is the  atomic number of the nucleus. This function should be defined by the user, an example is given in 
\verb|mdlIndep/dd_exp.c|.

\noindent
$\bullet$ \verb|DD_pval|(\verb|expCode|, $f_v$,\&expName)\\
$\bullet$  \verb|DD_factor|(\verb|expCode|, $\alpha$, $f_v$,\&expName)\\     
These functions are similar to  \verb|DD_pvalCS| and \verb|DD_factorCS| described above but use the  cross section calculated from the DM model under consideration in 
micrOMEGAs.  The necessary corrections for a light mediator are implemented automatically, these functions do not use {\tt dNdEFact}. \\

\noindent
The routines described above require  SD form factors when considering SD limits, by default they use the same  form factors as each  experiment. The  SD form factors can be replaced using the command \\
\noindent
$\bullet$ \verb| setSpinDepFF(ExperimentID, setID)|\\
where the choice for \verb|ExperimentID| is given at the beginning of this section, and \verb|setID| can be \\
 \verb|  EFT  -|  corresponding to  the form factors in ~\cite{Fitzpatrick:2012ix}, Eq.~\ref{eq:sdff_fitzpatrick} \\
  \verb|  SHELL -|  corresponding to  the average form factors in ~\cite{Klos:2013rwa}, Eq.~\ref{eq:middle}\\
  \verb|  SHELLm -|  corresponding to the  minimal form factor of  ~\cite{Klos:2013rwa}, Eq.~\ref{pm}.\\
See below.

\subsection{Spin-dependent form factors}  
\label{SDmin}
The  minimal and maximal values for the SD form factors, $S_{00}(q),S_{01}(q),S_{11}(q)$ , are computed in  Ref.\cite{Klos:2013rwa} within the shell model.
The flag  {\tt spinDepFlag=SHELL} corresponds to the average 
\begin{equation}
\label{eq:middle}
   S_{ab}=(S_{ab}^{min} +S_{ab}^{max})/2
\end{equation}
which are obtained from  the minimum and maximum fitted values  in Table VI in ~\cite{Klos:2013rwa}. 

Note that $S_{ab}^{min}$ often lead to a negative value for the subdominant component to the form factor. Since this has no physical meaning, to define the form factors that lead to
the most robust exclusion we rather use the minimum value of  the proton-only, $S_p^{min}$,   and neutron-only, $S_n^{min}$,  form factors also given in ~\cite{Klos:2013rwa}. These  correspond to the minimal form factor for the case when only one type of interaction (with proton or neutron) is included. With this we construct  the nucleus form factors 
\begin{eqnarray}
\label{pm}
  S_{00}&=& \frac{1}{4} \left(S_p^{min} +S_n^{min} \pm 2 \sqrt{S_p^{min} S_n^{min}}\right)\\
\nonumber
  S_{11}&=& \frac{1}{4}  \left(S_p^{min} +S_n^{min} \mp 2 \sqrt{S_p^{min} S_n^{min}}\right)\\
\nonumber
  S_{01}&=& \frac{1}{2}  (S_p^{min}-S_n^{min}) 
\end{eqnarray} 
The sign in Eq.\ref{pm} is chosen  to reproduce the  ratio $S_{00}(0)/{S}_{11}(0)$ for the central value of the form factors in Ref.~\cite{Klos:2013rwa}.

\subsection{Velocity distribution}
\label{veloAppendix} 

Ignoring the  direction of motion of DM particles and  the small effect of DM acceleration by the gravitational field of the Sun,
the DM velocity distribution in the vicinity  of the direct detection experiment is given by
 
$$f(\mbox{v})=  \int\limits_{|\vec{v}|<\rm{vEsc}} d^3\vec{v} F_G(\vec{v} -\vec{v}_{Earth})
\delta(\mbox{v} -|\vec{v}|) 
$$
where $F_G$ is the DM velocity distribution in the  frame, of the galaxy, $\vec{v}_{Earth}$ is the velocity of the
Earth in the Galaxy and {\tt vEsc} is the maximal velocity in our Galaxy due to its finite gravitationnal
potential. {\tt vEsc} and {\tt vEarth=}$|\vec{v}_{Earth}|$ are global
parameters of micrOMEGAs. 

The velocity distributions that are available  in \micro~ are the following

\noindent
$\bullet$ \verb|Maxwell(v)| \\
returns   
$$F_G^M(\mbox{v})=  c_{\rm{norm}} \frac{1}{(2\pi \rm{vRot}^2)^{3/2}} \exp\left(
-\frac{(\vec{v})^2}{\rm{vRot}^2}\right)\theta(\mbox{vEsc} -|\vec{v}|)
$$
which corresponds to the isothermal model. Here  {\tt vRot} is the orbital velocity of stars in the Milky Way, it is also a  global 
parameter of micrOMEGAs. $c_{\tt norm}$ is the normalization factor,

$$
c_{\tt norm}^{-1}=  {\mbox{erf}}\left(\frac{\mbox{vEsc}}{\rm{vRot}} \right) -\frac{2}{\sqrt{\pi}} \frac{\mbox{vEsc}}{\rm{vRot}} \exp\left( -\frac{\mbox{vEsc}^2}{\rm{vRot}^2}\right)
$$  

\noindent
$\bullet$ \verb|SHMpp(v)|\\
returns the  velocity distribution {\tt SHM++}  proposed in ~\cite{Evans:2018bqy}.
\begin{equation}
F_G(\vec{v}) = (1-\eta) F_G^M(\mbox{v}) +\eta F_G^S(\mbox{v})
\end{equation}

This distribution consists of two components. The first, $F_G^M(\vec{v})$,   
is  the standard Maxwell velocity distribution described above.   The second component  is the velocity distribution from the {\it Gaia} sausage ~\cite{Belokurov_2018,Myeong:2018kfh}, it  is not spherically symmetric and is defined by the
anisotropy parameter $\beta$ with
\begin{equation}
 F_G^S(\vec{v})=   \frac{c_{\rm{norm}} }{(2\pi)^{3/2} \Delta v_r \Delta v_\theta\Delta v_\phi}
  \exp\left( -\left( \frac{v_r}{\Delta v_r}\right)^2 
 -\left( \frac{v_\theta}{\Delta v_\theta}\right)^2  -\left( \frac{v_\phi}{\Delta v_\phi}\right)^2   \right)
\theta(\mbox{vEsc} -|\vec{v}|)
\end{equation}
where  
\begin{equation}
\Delta v_r= \frac{\rm{vRot}}{\sqrt{1-\frac{2}{3}\beta}}\;,\;\; \Delta v_\phi=\Delta v_\theta=\frac{\rm{vRot} \sqrt{1-\beta}}{\sqrt{1-\frac{2}{3}\beta}} 
\end{equation}  
and 
$$
c_{\tt norm}^{-1}= {\mbox{erf}}\left(\frac{\mbox{vEsc}}{\rm{vRot}} \right) - \left( \frac{1-\beta}{\beta} \right)^{1/2}   \exp\left( -\frac{\mbox{vEsc}^2}{\rm{vRot}^2}\right) 
{\mbox{erfi}}\left( \frac{\mbox{vEsc}}{\rm{vRot}}\frac{\beta^{1/2}}{(1-\beta)^{1/2}}        \right)
$$  
where ${\mbox{erfi}}$ is the imaginary error function.
  
The central values  and uncertainties   of   the {\tt SHM++} parameters are  
\begin{eqnarray}
\nonumber 
\textsf{rhoDM} &=&0.55\pm 0.17 ~{\rm GeV}/{\rm cm}^3\\
\nonumber
\textsf{vRot}&=& 233\pm 3 ~{\rm km/s}\\
\nonumber
\textsf{vEsc}&=& 580\pm 63~ {\rm km/s}\\
\nonumber
 \beta=\textsf{betaSHMpp} &=&0.9\pm 0.05\\
\eta= \textsf{etaSHMpp} &=& 0.2\pm 0.1 
\label{eq:SHMpp} 
\end{eqnarray}
\noindent
Note that these central values for the global parameters, {\tt vRot, vEsc} and {\tt rhoDM}   
are different from  the  ones in Eq.\ref{velo_std}.


\bibliography{xenon}

\providecommand{\href}[2]{#2}\begingroup\raggedright\begin{thebibliography}{10}

\bibitem{Aprile:2018dbl}
{\bfseries XENON} Collaboration, E.~Aprile {\em et~al.}, ``{Dark Matter Search
  Results from a One Tonne$\times$Year Exposure of XENON1T},''
\href{http://arxiv.org/abs/1805.12562}{{\ttfamily arXiv:1805.12562
  [astro-ph.CO]}}.

\bibitem{Aprile:2019dbj}
{\bfseries XENON} Collaboration, E.~Aprile {\em et~al.}, ``{Constraining the
  spin-dependent WIMP-nucleon cross sections with XENON1T},''
  \href{http://dx.doi.org/10.1103/PhysRevLett.122.141301}{{\em Phys. Rev.
  Lett.} {\bfseries 122} no.~14, (2019) 141301},
\href{http://arxiv.org/abs/1902.03234}{{\ttfamily arXiv:1902.03234
  [astro-ph.CO]}}.

\bibitem{Agnes:2018ves}
{\bfseries DarkSide} Collaboration, P.~Agnes {\em et~al.}, ``{Low-Mass Dark
  Matter Search with the DarkSide-50 Experiment},''
  \href{http://dx.doi.org/10.1103/PhysRevLett.121.081307}{{\em Phys. Rev.
  Lett.} {\bfseries 121} no.~8, (2018) 081307},
\href{http://arxiv.org/abs/1802.06994}{{\ttfamily arXiv:1802.06994
  [astro-ph.HE]}}.

\bibitem{Amole:2017dex}
{\bfseries PICO} Collaboration, C.~Amole {\em et~al.}, ``{Dark Matter Search
  Results from the PICO-60 C$_3$F$_8$ Bubble Chamber},''
  \href{http://dx.doi.org/10.1103/PhysRevLett.118.251301}{{\em Phys. Rev.
  Lett.} {\bfseries 118} no.~25, (2017) 251301},
  \href{http://arxiv.org/abs/1702.07666}{{\ttfamily arXiv:1702.07666
  [astro-ph.CO]}}.

\bibitem{Amole:2019fdf}
{\bfseries PICO} Collaboration, C.~Amole {\em et~al.}, ``{Dark Matter Search
  Results from the Complete Exposure of the PICO-60 C$_3$F$_8$ Bubble
  Chamber},'' \href{http://dx.doi.org/10.1103/PhysRevD.100.022001}{{\em Phys.
  Rev.} {\bfseries D100} no.~2, (2019) 022001},
\href{http://arxiv.org/abs/1902.04031}{{\ttfamily arXiv:1902.04031
  [astro-ph.CO]}}.

\bibitem{Petricca:2017zdp}
{\bfseries CRESST} Collaboration, F.~Petricca {\em et~al.}, ``{First results on
  low-mass dark matter from the CRESST-III experiment},''
\href{http://arxiv.org/abs/1711.07692}{{\ttfamily arXiv:1711.07692
  [astro-ph.CO]}}.

\bibitem{Xia:2018qgs}
{\bfseries PandaX-II} Collaboration, J.~Xia {\em et~al.}, ``{PandaX-II
  Constraints on Spin-Dependent WIMP-Nucleon Effective Interactions},''
  \href{http://dx.doi.org/10.1016/j.physletb.2019.02.043}{{\em Phys. Lett.}
  {\bfseries B792} (2019) 193--198},
\href{http://arxiv.org/abs/1807.01936}{{\ttfamily arXiv:1807.01936 [hep-ex]}}.

\bibitem{Agnese:2018gze}
{\bfseries SuperCDMS} Collaboration, R.~Agnese {\em et~al.}, ``{Search for
  Low-Mass Dark Matter with CDMSlite Using a Profile Likelihood Fit},''
  \href{http://dx.doi.org/10.1103/PhysRevD.99.062001}{{\em Phys. Rev.}
  {\bfseries D99} no.~6, (2019) 062001},
\href{http://arxiv.org/abs/1808.09098}{{\ttfamily arXiv:1808.09098
  [astro-ph.CO]}}.

\bibitem{Aprile:2019xxb}
{\bfseries XENON} Collaboration, E.~Aprile {\em et~al.}, ``{Light Dark Matter
  Search with Ionization Signals in XENON1T},''
  \href{http://dx.doi.org/10.1103/PhysRevLett.123.251801}{{\em Phys. Rev.
  Lett.} {\bfseries 123} no.~25, (2019) 251801},
\href{http://arxiv.org/abs/1907.11485}{{\ttfamily arXiv:1907.11485 [hep-ex]}}.

\bibitem{Abdelhameed:2019hmk}
{\bfseries CRESST} Collaboration, A.~H. Abdelhameed {\em et~al.}, ``{First
  results from the CRESST-III low-mass dark matter program},''
  \href{http://dx.doi.org/10.1103/PhysRevD.100.102002}{{\em Phys. Rev.}
  {\bfseries D100} no.~10, (2019) 102002},
\href{http://arxiv.org/abs/1904.00498}{{\ttfamily arXiv:1904.00498
  [astro-ph.CO]}}.

\bibitem{Agnese:2015nto}
{\bfseries SuperCDMS} Collaboration, R.~Agnese {\em et~al.}, ``{New Results
  from the Search for Low-Mass Weakly Interacting Massive Particles with the
  CDMS Low Ionization Threshold Experiment},''
  \href{http://dx.doi.org/10.1103/PhysRevLett.116.071301}{{\em Phys. Rev.
  Lett.} {\bfseries 116} no.~7, (2016) 071301},
\href{http://arxiv.org/abs/1509.02448}{{\ttfamily arXiv:1509.02448
  [astro-ph.CO]}}.

\bibitem{Aprile:2019jmx}
{\bfseries XENON} Collaboration, E.~Aprile {\em et~al.}, ``{Search for Light
  Dark Matter Interactions Enhanced by the Migdal Effect or Bremsstrahlung in
  XENON1T},'' \href{http://dx.doi.org/10.1103/PhysRevLett.123.241803}{{\em
  Phys. Rev. Lett.} {\bfseries 123} no.~24, (2019) 241803},
\href{http://arxiv.org/abs/1907.12771}{{\ttfamily arXiv:1907.12771 [hep-ex]}}.

\bibitem{Essig:2011nj}
R.~Essig, J.~Mardon, and T.~Volansky, ``{Direct Detection of Sub-GeV Dark
  Matter},'' \href{http://dx.doi.org/10.1103/PhysRevD.85.076007}{{\em Phys.
  Rev.} {\bfseries D85} (2012) 076007},
\href{http://arxiv.org/abs/1108.5383}{{\ttfamily arXiv:1108.5383 [hep-ph]}}.

\bibitem{Agnese:2018col}
{\bfseries SuperCDMS} Collaboration, R.~Agnese {\em et~al.}, ``{First Dark
  Matter Constraints from a SuperCDMS Single-Charge Sensitive Detector},''
  \href{http://dx.doi.org/10.1103/PhysRevLett.122.069901,
  10.1103/PhysRevLett.121.051301}{{\em Phys. Rev. Lett.} {\bfseries 121} no.~5,
  (2018) 051301}, \href{http://arxiv.org/abs/1804.10697}{{\ttfamily
  arXiv:1804.10697 [hep-ex]}}.
[erratum: Phys. Rev. Lett.122,no.6,069901(2019)].

\bibitem{Abramoff:2019dfb}
{\bfseries SENSEI} Collaboration, O.~Abramoff {\em et~al.}, ``{SENSEI:
  Direct-Detection Constraints on Sub-GeV Dark Matter from a Shallow
  Underground Run Using a Prototype Skipper-CCD},''
  \href{http://dx.doi.org/10.1103/PhysRevLett.122.161801}{{\em Phys. Rev.
  Lett.} {\bfseries 122} no.~16, (2019) 161801},
\href{http://arxiv.org/abs/1901.10478}{{\ttfamily arXiv:1901.10478 [hep-ex]}}.

\bibitem{Aguilar-Arevalo:2019wdi}
{\bfseries DAMIC} Collaboration, A.~Aguilar-Arevalo {\em et~al.},
  ``{Constraints on Light Dark Matter Particles Interacting with Electrons from
  DAMIC at SNOLAB},''
  \href{http://dx.doi.org/10.1103/PhysRevLett.123.181802}{{\em Phys. Rev.
  Lett.} {\bfseries 123} no.~18, (2019) 181802},
\href{http://arxiv.org/abs/1907.12628}{{\ttfamily arXiv:1907.12628
  [astro-ph.CO]}}.

\bibitem{Arnaud:2020svb}
{\bfseries EDELWEISS} Collaboration, Q.~Arnaud {\em et~al.}, ``{First
  germanium-based constraints on sub-MeV Dark Matter with the EDELWEISS
  experiment},''
\href{http://arxiv.org/abs/2003.01046}{{\ttfamily arXiv:2003.01046
  [astro-ph.GA]}}.

\bibitem{Aprile:2020tmw}
{\bfseries XENON} Collaboration, E.~Aprile {\em et~al.}, ``{Excess electronic
  recoil events in XENON1T},''
  \href{http://dx.doi.org/10.1103/PhysRevD.102.072004}{{\em Phys. Rev. D}
  {\bfseries 102} no.~7, (2020) 072004},
  \href{http://arxiv.org/abs/2006.09721}{{\ttfamily arXiv:2006.09721
  [hep-ex]}}.

\bibitem{Buckley:2009in}
M.~R. Buckley and P.~J. Fox, ``{Dark Matter Self-Interactions and Light Force
  Carriers},'' \href{http://dx.doi.org/10.1103/PhysRevD.81.083522}{{\em Phys.
  Rev.} {\bfseries D81} (2010) 083522},
\href{http://arxiv.org/abs/0911.3898}{{\ttfamily arXiv:0911.3898 [hep-ph]}}.

\bibitem{Bringmann:2016din}
T.~Bringmann, F.~Kahlhoefer, K.~Schmidt-Hoberg, and P.~Walia, ``{Strong
  constraints on self-interacting dark matter with light mediators},''
  \href{http://dx.doi.org/10.1103/PhysRevLett.118.141802}{{\em Phys. Rev.
  Lett.} {\bfseries 118} no.~14, (2017) 141802},
\href{http://arxiv.org/abs/1612.00845}{{\ttfamily arXiv:1612.00845 [hep-ph]}}.

\bibitem{Kahlhoefer:2017ddj}
F.~Kahlhoefer, S.~Kulkarni, and S.~Wild, ``{Exploring light mediators with
  low-threshold direct detection experiments},''
  \href{http://dx.doi.org/10.1088/1475-7516/2017/11/016}{{\em JCAP} {\bfseries
  1711} no.~11, (2017) 016},
\href{http://arxiv.org/abs/1707.08571}{{\ttfamily arXiv:1707.08571 [hep-ph]}}.

\bibitem{Aarssen:2012fx}
L.~G. van~den Aarssen, T.~Bringmann, and C.~Pfrommer, ``{Is dark matter with
  long-range interactions a solution to all small-scale problems of $\Lambda$
  CDM cosmology?},''
  \href{http://dx.doi.org/10.1103/PhysRevLett.109.231301}{{\em Phys. Rev.
  Lett.} {\bfseries 109} (2012) 231301},
\href{http://arxiv.org/abs/1205.5809}{{\ttfamily arXiv:1205.5809
  [astro-ph.CO]}}.

\bibitem{Tulin:2013teo}
S.~Tulin, H.-B. Yu, and K.~M. Zurek, ``{Beyond Collisionless Dark Matter:
  Particle Physics Dynamics for Dark Matter Halo Structure},''
  \href{http://dx.doi.org/10.1103/PhysRevD.87.115007}{{\em Phys. Rev.}
  {\bfseries D87} no.~11, (2013) 115007},
\href{http://arxiv.org/abs/1302.3898}{{\ttfamily arXiv:1302.3898 [hep-ph]}}.

\bibitem{Kaplinghat:2015aga}
M.~Kaplinghat, S.~Tulin, and H.-B. Yu, ``{Dark Matter Halos as Particle
  Colliders: Unified Solution to Small-Scale Structure Puzzles from Dwarfs to
  Clusters},'' \href{http://dx.doi.org/10.1103/PhysRevLett.116.041302}{{\em
  Phys. Rev. Lett.} {\bfseries 116} no.~4, (2016) 041302},
\href{http://arxiv.org/abs/1508.03339}{{\ttfamily arXiv:1508.03339
  [astro-ph.CO]}}.

\bibitem{Hambye:2018dpi}
T.~Hambye, M.~H.~G. Tytgat, J.~Vandecasteele, and L.~Vanderheyden, ``{Dark
  matter direct detection is testing freeze-in},''
  \href{http://dx.doi.org/10.1103/PhysRevD.98.075017}{{\em Phys. Rev.}
  {\bfseries D98} no.~7, (2018) 075017},
\href{http://arxiv.org/abs/1807.05022}{{\ttfamily arXiv:1807.05022 [hep-ph]}}.

\bibitem{Cui:2017nnn}
{\bfseries PandaX-II} Collaboration, X.~Cui {\em et~al.}, ``{Dark Matter
  Results From 54-Ton-Day Exposure of PandaX-II Experiment},''
  \href{http://dx.doi.org/10.1103/PhysRevLett.119.181302}{{\em Phys. Rev.
  Lett.} {\bfseries 119} no.~18, (2017) 181302},
\href{http://arxiv.org/abs/1708.06917}{{\ttfamily arXiv:1708.06917
  [astro-ph.CO]}}.

\bibitem{Belanger:2008sj}
G.~B\'elanger, F.~Boudjema, A.~Pukhov, and A.~Semenov, ``{Dark matter direct
  detection rate in a generic model with micrOMEGAs 2.2},''
  \href{http://dx.doi.org/10.1016/j.cpc.2008.11.019}{{\em Comput. Phys.
  Commun.} {\bfseries 180} (2009) 747--767},
\href{http://arxiv.org/abs/0803.2360}{{\ttfamily arXiv:0803.2360 [hep-ph]}}.

\bibitem{Belanger:2018mqt}
G.~B\'elanger, F.~Boudjema, A.~Goudelis, A.~Pukhov, and B.~Zaldivar,
  ``{micrOMEGAs5.0 : Freeze-in},''
  \href{http://dx.doi.org/10.1016/j.cpc.2018.04.027}{{\em Comput. Phys.
  Commun.} {\bfseries 231} (2018) 173--186},
\href{http://arxiv.org/abs/1801.03509}{{\ttfamily arXiv:1801.03509 [hep-ph]}}.

\bibitem{Workgroup:2017lvb}
{\bfseries The GAMBIT Dark Matter Workgroup} Collaboration, T.~Bringmann {\em
  et~al.}, ``{DarkBit: A GAMBIT module for computing dark matter observables
  and likelihoods},''
  \href{http://dx.doi.org/10.1140/epjc/s10052-017-5155-4}{{\em Eur. Phys. J.}
  {\bfseries C77} no.~12, (2017) 831},
\href{http://arxiv.org/abs/1705.07920}{{\ttfamily arXiv:1705.07920 [hep-ph]}}.

\bibitem{Athron:2018ipf}
P.~Athron, J.~M. Cornell, F.~Kahlhoefer, J.~Mckay, P.~Scott, and S.~Wild,
  ``{Impact of vacuum stability, perturbativity and XENON1T on global fits of
  $\mathbb {Z}_2$ and $\mathbb {Z}_3$ scalar singlet dark matter},''
  \href{http://dx.doi.org/10.1140/epjc/s10052-018-6314-y}{{\em Eur. Phys. J. C}
  {\bfseries 78} no.~10, (2018) 830},
  \href{http://arxiv.org/abs/1806.11281}{{\ttfamily arXiv:1806.11281
  [hep-ph]}}.

\bibitem{Athron:2018hpc}
{\bfseries GAMBIT} Collaboration, P.~Athron {\em et~al.}, ``{Global analyses of
  Higgs portal singlet dark matter models using GAMBIT},''
  \href{http://dx.doi.org/10.1140/epjc/s10052-018-6513-6}{{\em Eur. Phys. J. C}
  {\bfseries 79} no.~1, (2019) 38},
  \href{http://arxiv.org/abs/1808.10465}{{\ttfamily arXiv:1808.10465
  [hep-ph]}}.

\bibitem{Arbey:2018msw}
A.~Arbey, F.~Mahmoudi, and G.~Robbins, ``{SuperIso Relic v4: A program for
  calculating dark matter and flavour physics observables in Supersymmetry},''
  \href{http://dx.doi.org/10.1016/j.cpc.2019.01.014}{{\em Comput. Phys.
  Commun.} {\bfseries 239} (2019) 238--264},
\href{http://arxiv.org/abs/1806.11489}{{\ttfamily arXiv:1806.11489 [hep-ph]}}.

\bibitem{Aprile:2017aas}
{\bfseries XENON} Collaboration, E.~Aprile {\em et~al.}, ``{Effective field
  theory search for high-energy nuclear recoils using the XENON100 dark matter
  detector},'' \href{http://dx.doi.org/10.1103/PhysRevD.96.042004}{{\em Phys.
  Rev.} {\bfseries D96} no.~4, (2017) 042004},
\href{http://arxiv.org/abs/1705.02614}{{\ttfamily arXiv:1705.02614
  [astro-ph.CO]}}.

\bibitem{Angloher:2018fcs}
{\bfseries CRESST} Collaboration, G.~Angloher {\em et~al.}, ``{Limits on Dark
  Matter Effective Field Theory Parameters with CRESST-II},''
  \href{http://dx.doi.org/10.1140/epjc/s10052-018-6523-4}{{\em Eur. Phys. J.}
  {\bfseries C79} no.~1, (2019) 43},
\href{http://arxiv.org/abs/1809.03753}{{\ttfamily arXiv:1809.03753 [hep-ph]}}.

\bibitem{Bozorgnia:2018jep}
N.~Bozorgnia, D.~G. Cerdeño, A.~Cheek, and B.~Penning, ``{Opening the energy
  window on direct dark matter detection},''
  \href{http://dx.doi.org/10.1088/1475-7516/2018/12/013}{{\em JCAP} {\bfseries
  1812} no.~12, (2018) 013},
\href{http://arxiv.org/abs/1810.05576}{{\ttfamily arXiv:1810.05576 [hep-ph]}}.

\bibitem{Lewin:1995rx}
J.~D. Lewin and P.~F. Smith, ``{Review of mathematics, numerical factors, and
  corrections for dark matter experiments based on elastic nuclear recoil},''
\href{http://dx.doi.org/10.1016/S0927-6505(96)00047-3}{{\em Astropart. Phys.}
  {\bfseries 6} (1996) 87--112}.

\bibitem{Bednyakov:2006ux}
V.~A. Bednyakov and F.~Simkovic, ``{Nuclear spin structure in dark matter
  search: The Finite momentum transfer limit},''
  \href{http://dx.doi.org/10.1134/S1063779606070057}{{\em Phys. Part. Nucl.}
  {\bfseries 37} (2006) S106--S128},
\href{http://arxiv.org/abs/hep-ph/0608097}{{\ttfamily arXiv:hep-ph/0608097
  [hep-ph]}}.

\bibitem{Klos:2013rwa}
P.~Klos, J.~Men\'endez, D.~Gazit, and A.~Schwenk, ``{Large-scale nuclear
  structure calculations for spin-dependent WIMP scattering with chiral
  effective field theory currents},''
  \href{http://dx.doi.org/10.1103/PhysRevD.89.029901,
  10.1103/PhysRevD.88.083516}{{\em Phys. Rev.} {\bfseries D88} no.~8, (2013)
  083516}, \href{http://arxiv.org/abs/1304.7684}{{\ttfamily arXiv:1304.7684
  [nucl-th]}}.
[Erratum: Phys. Rev.D89,no.2,029901(2014)].

\bibitem{Fitzpatrick:2012ix}
A.~L. Fitzpatrick, W.~Haxton, E.~Katz, N.~Lubbers, and Y.~Xu, ``{The Effective
  Field Theory of Dark Matter Direct Detection},''
  \href{http://dx.doi.org/10.1088/1475-7516/2013/02/004}{{\em JCAP} {\bfseries
  1302} (2013) 004},
\href{http://arxiv.org/abs/1203.3542}{{\ttfamily arXiv:1203.3542 [hep-ph]}}.

\bibitem{Aprile:2019dme}
{\bfseries XENON} Collaboration, E.~Aprile {\em et~al.}, ``{XENON1T dark matter
  data analysis: Signal and background models and statistical inference},''
  \href{http://dx.doi.org/10.1103/PhysRevD.99.112009}{{\em Phys. Rev. D}
  {\bfseries 99} no.~11, (2019) 112009},
  \href{http://arxiv.org/abs/1902.11297}{{\ttfamily arXiv:1902.11297
  [physics.ins-det]}}.

\bibitem{Yellin:2008da}
S.~Yellin, ``{Extending the optimum interval method},''
  \href{http://arxiv.org/abs/0709.2701}{{\ttfamily arXiv:0709.2701
  [physics.data-an]}}.
\url{http://cdms.stanford.edu/Upperlimit/}.

\bibitem{Cowan:2010js}
G.~Cowan, K.~Cranmer, E.~Gross, and O.~Vitells, ``{Asymptotic formulae for
  likelihood-based tests of new physics},''
  \href{http://dx.doi.org/10.1140/epjc/s10052-011-1554-0,
  10.1140/epjc/s10052-013-2501-z}{{\em Eur. Phys. J.} {\bfseries C71} (2011)
  1554}, \href{http://arxiv.org/abs/1007.1727}{{\ttfamily arXiv:1007.1727
  [physics.data-an]}}.
[Erratum: Eur. Phys. J.C73,2501(2013)].

\bibitem{Bezrukov:2010qa}
F.~Bezrukov, F.~Kahlhoefer, and M.~Lindner, ``{Interplay between scintillation
  and ionization in liquid xenon Dark Matter searches},''
  \href{http://dx.doi.org/10.1016/j.astropartphys.2011.06.008}{{\em Astropart.
  Phys.} {\bfseries 35} (2011) 119--127},
\href{http://arxiv.org/abs/1011.3990}{{\ttfamily arXiv:1011.3990
  [astro-ph.IM]}}.

\bibitem{Bernreuther:2019pfb}
E.~Bernreuther, F.~Kahlhoefer, M.~Krämer, and P.~Tunney, ``{Strongly
  interacting dark sectors in the early Universe and at the LHC through a
  simplified portal},'' \href{http://dx.doi.org/10.1007/JHEP01(2020)162}{{\em
  JHEP} {\bfseries 01} (2020) 162},
  \href{http://arxiv.org/abs/1907.04346}{{\ttfamily arXiv:1907.04346
  [hep-ph]}}.

\bibitem{Feldman:1997qc}
G.~J. Feldman and R.~D. Cousins, ``{A Unified approach to the classical
  statistical analysis of small signals},''
  \href{http://dx.doi.org/10.1103/PhysRevD.57.3873}{{\em Phys. Rev.} {\bfseries
  D57} (1998) 3873--3889},
\href{http://arxiv.org/abs/physics/9711021}{{\ttfamily arXiv:physics/9711021
  [physics.data-an]}}.

\bibitem{Yellin:2002xd}
S.~Yellin, ``{Finding an upper limit in the presence of unknown background},''
  \href{http://dx.doi.org/10.1103/PhysRevD.66.032005}{{\em Phys. Rev.}
  {\bfseries D66} (2002) 032005},
\href{http://arxiv.org/abs/physics/0203002}{{\ttfamily arXiv:physics/0203002
  [physics]}}.

\bibitem{Abdelhameed:2019mac}
{\bfseries CRESST} Collaboration, A.~H. Abdelhameed {\em et~al.},
  ``{Description of CRESST-III Data},''
\href{http://arxiv.org/abs/1905.07335}{{\ttfamily arXiv:1905.07335
  [astro-ph.CO]}}.

\bibitem{Haas:2014dda}
A.~Haas, C.~S. Hill, E.~Izaguirre, and I.~Yavin, ``{Looking for milli-charged
  particles with a new experiment at the LHC},''
  \href{http://dx.doi.org/10.1016/j.physletb.2015.04.062}{{\em Phys. Lett.}
  {\bfseries B746} (2015) 117--120},
\href{http://arxiv.org/abs/1410.6816}{{\ttfamily arXiv:1410.6816 [hep-ph]}}.

\bibitem{Landau1981Quantum}
L.~D. Landau and L.~M. Lifshitz, {\em Quantum Mechanics Non-Relativistic
  Theory, Third Edition: Volume 3}.
\newblock Butterworth-Heinemann, 3~ed., Jan., 1981.
\newblock \url{http://www.worldcat.org/isbn/0750635398}.

\bibitem{Tanabashi:2018oca}
{\bfseries ParticleDataGroup} Collaboration, M.~Tanabashi {\em et~al.},
  ``{Review of Particle Physics},''
\href{http://dx.doi.org/10.1103/PhysRevD.98.030001}{{\em Phys. Rev.} {\bfseries
  D98} no.~3, (2018) 030001}.

\bibitem{Salucci:2010qr}
P.~Salucci, F.~Nesti, G.~Gentile, and C.~F. Martins, ``{The dark matter density
  at the Sun's location},''
  \href{http://dx.doi.org/10.1051/0004-6361/201014385}{{\em Astron. Astrophys.}
  {\bfseries 523} (2010) A83},
\href{http://arxiv.org/abs/1003.3101}{{\ttfamily arXiv:1003.3101
  [astro-ph.GA]}}.

\bibitem{Green:2017odb}
A.~M. Green, ``{Astrophysical uncertainties on the local dark matter
  distribution and direct detection experiments},''
  \href{http://dx.doi.org/10.1088/1361-6471/aa7819}{{\em J. Phys.} {\bfseries
  G44} no.~8, (2017) 084001},
\href{http://arxiv.org/abs/1703.10102}{{\ttfamily arXiv:1703.10102
  [astro-ph.CO]}}.

\bibitem{Monari_2018}
G.~Monari, B.~Famaey, I.~Carrillo, T.~Piffl, M.~Steinmetz, R.~F.~G. Wyse,
  F.~Anders, C.~Chiappini, and K.~Janßen, ``The escape speed curve of the
  galaxy obtained fromgaiadr2 implies a heavy milky way,''
  \href{http://dx.doi.org/10.1051/0004-6361/201833748}{{\em Astronomy \&
  Astrophysics} {\bfseries 616} (Aug, 2018) L9}.
  \url{http://dx.doi.org/10.1051/0004-6361/201833748}.

\bibitem{Wu:2019nhd}
Y.~Wu, K.~Freese, C.~Kelso, P.~Stengel, and M.~Valluri, ``{Uncertainties in
  Direct Dark Matter Detection in Light of Gaia's Escape Velocity
  Measurements},'' \href{http://dx.doi.org/10.1088/1475-7516/2019/10/034}{{\em
  JCAP} {\bfseries 1910} no.~10, (2019) 034},
\href{http://arxiv.org/abs/1904.04781}{{\ttfamily arXiv:1904.04781 [hep-ph]}}.

\bibitem{Evans:2018bqy}
N.~W. Evans, C.~A.~J. O'Hare, and C.~McCabe, ``{Refinement of the standard halo
  model for dark matter searches in light of the Gaia Sausage},''
  \href{http://dx.doi.org/10.1103/PhysRevD.99.023012}{{\em Phys. Rev.}
  {\bfseries D99} no.~2, (2019) 023012},
\href{http://arxiv.org/abs/1810.11468}{{\ttfamily arXiv:1810.11468
  [astro-ph.GA]}}.

\bibitem{Belokurov_2018}
V.~Belokurov, D.~Erkal, N.~W. Evans, S.~E. Koposov, and A.~J. Deason,
  ``Co-formation of the disc and the stellar halo,''
  \href{http://dx.doi.org/10.1093/mnras/sty982}{{\em Monthly Notices of the
  Royal Astronomical Society} {\bfseries 478} no.~1, (Jun, 2018) 611–619}.
  \url{http://dx.doi.org/10.1093/mnras/sty982}.

\bibitem{Myeong:2018kfh}
G.~C. Myeong, N.~W. Evans, V.~Belokurov, J.~L. Sanders, and S.~E. Koposov,
  ``{The Sausage Globular Clusters},''
  \href{http://dx.doi.org/10.3847/2041-8213/aad7f7}{{\em Astrophys. J.}
  {\bfseries 863} no.~2, (2018) L28},
\href{http://arxiv.org/abs/1805.00453}{{\ttfamily arXiv:1805.00453
  [astro-ph.GA]}}.

\end{thebibliography}\endgroup

\end{document}